\documentclass[journal]{IEEEtran}

\usepackage{amsmath}
\usepackage{amssymb}
\usepackage{amsfonts}
\usepackage{epsfig}
\usepackage{cite}
\usepackage{calc}
\usepackage{color}

\title{Explicit Space-Time Codes Achieving The Diversity-Multiplexing Gain Tradeoff}
\author{Petros Elia, K. Raj Kumar, Sameer A. Pawar \\
P. Vijay Kumar and Hsiao-feng Lu
\thanks{Petros Elia and P. Vijay Kumar are with the Department of EE-Systems,
University of Southern California, Los Angeles, CA 90089 ({\tt
\{elia,vijayk\}@usc.edu}).  K. Raj Kumar and Sameer A. Pawar are
with the Department of Electrical Communication Engineering of the
Indian Institute of Science, Bangalore, 560 012 ({\tt
\{raj,sameerp\}@ece.iisc.ernet.in}). This work was carried out
while P. Vijay Kumar was on leave of absence at the Indian
Institute of Science, Bangalore.  Hsiao-feng (Francis) Lu is with
the Dept. of Comm. Engineering, National Chung-cheng University,
160 San-Hsing, Min-Hsiung, Chia-Yi 621, Taiwan, R.O.C. ({\tt
francis@ccu.edu.tw}).}
\thanks{This research is supported in part by NSF-ITR CCR-0326628, in part by
the DRDO-IISc Program on Advanced Research in Mathematical
Engineering and in part from Grant NSC 93-2218-E-194-012.} }

\newcommand{\Z}{{\mathbb Z}}

\renewcommand{\frak}{\mathfrak}

\newtheorem{thm}{Theorem}
\newtheorem{prop}[thm]{Proposition}
\newtheorem{lem}[thm]{Lemma}
\newtheorem{cor}[thm]{Corollary}

\newtheorem{ex}{{\em Example}}

\newcommand{\beq}{\begin{equation}}
\newcommand{\eeq}{\end{equation}}

\newcommand{\bea}{\begin{eqnarray}}
\newcommand{\eea}{\end{eqnarray}}
\newcommand{\bean}{\begin{eqnarray*}}
\newcommand{\eean}{\end{eqnarray*}}

\newcommand{\bit}{\begin{itemize}}
\newcommand{\eit}{\end{itemize}}
\newcommand{\ben}{\begin{enumerate}}
\newcommand{\een}{\end{enumerate}}

\begin{document}

\maketitle\thispagestyle{empty}

%\tableofcontents
%\newpage

\bibliographystyle{ieeetran}
\begin{abstract}

A recent result of Zheng and Tse states that over a quasi-static
channel, there exists a fundamental tradeoff, referred to as the
diversity-multiplexing gain (D-MG) tradeoff, between the spatial
multiplexing gain and the diversity gain that can be
simultaneously achieved by a space-time (ST) block code. This tradeoff
is precisely known in the case of i.i.d. Rayleigh-fading, for $T
\geq n_t+n_r-1$ where $T$ is the number of time slots over which
coding takes place and $n_t,n_r$ are the number of transmit and
receive antennas respectively.  For $T < n_t+n_r-1$, only upper
and lower bounds on the D-MG tradeoff are available.

In this paper, we present a complete solution to the problem of
explicitly constructing D-MG optimal ST codes, i.e., codes that
achieve the D-MG tradeoff for any number of receive antennas.  We
do this by showing that for the square minimum-delay case when
$T=n_t=n$, cyclic-division-algebra (CDA) based ST codes having the
non-vanishing determinant property are D-MG optimal. While
constructions of such codes were previously known for restricted
values of $n$, we provide here a construction for such codes that
is valid for all $n$.

For the rectangular, $T > n_t$ case, we present two general
techniques for building D-MG-optimal rectangular ST codes from
their square counterparts. A byproduct of our results establishes
that the D-MG tradeoff for all $T\geq n_t$ is the same as that
previously known to hold for $T \geq n_t + n_r -1$.

\end{abstract}

\normalsize

%\newpage

%why cannot r exceed min {n_r,n_t} ?
%Xia_on optimal 2 by 2

\section{Introduction \label{sec:intro}}
Consider the quasi-static, Rayleigh fading, space-time (ST) MIMO
channel with quasi-static interval $T$, $n_t$ transmit and $n_r$
receive antennas. The $(n_r \times T)$ received signal matrix $Y$
is given by
\begin{equation}
Y \ = \ \theta H X + W \label{eq:model}
\end{equation}
 where $X$ is a $(n_t \times T)$ code matrix drawn from a ST code ${\cal X}$,
 %associated to the space-time (ST) code,
 $H$ the $(n_r \times n_t)$ channel matrix and $W$ represents additive noise.
The entries of $H$ and $W$ are assumed to be i.i.d., circularly
symmetric, complex Gaussian $\mathbb{C}\mathcal {N} (0,1)$ random
variables.  The real scalar $\theta$ ensures that the energy
constraint \bea \theta^2 ||X||_F^2 \ \leq \ T \ \text{SNR}, \ \
\text{ all $X \in {\cal X}$}, \label{eq:energy_condition_intro}
\eea
 is met. We set
\[ {\cal Z} \ = \ \{ \theta X \mid X \in {\cal X} \}\] and will
refer to ${\cal X},{\cal Z}$ as the unnormalized and normalized ST
codes \footnote{As pointed out by the reviewers, the collection of
matrices ${\cal X}$ could also be regarded as forming a signal
constellation in which case the term ST modulation might be more
appropriate.  However, we have retained the ST code label here to
be in keeping with the widespread usage of this terminology in the
literature.} respectively.

Multiple transmit and receive antennas have the potential of
increasing reliability of communication as well as permitting
communication at higher rates. These aspects are quantified by the
diversity and spatial multiplexing gains respectively. In a recent
landmark paper, Zheng and Tse \cite{ZheTse} showed that there is a
fundamental tradeoff explained below, between diversity and
multiplexing gain, referred to as the diversity-multiplexing gain
(D-MG) tradeoff.

The ergodic capacity\cite{Tel,Fos}, i.e., capacity averaged over
all realizations $H$ of the space-time channel model in
\eqref{eq:model} is given by
\[
C \ = \ \mathbb{E} \{ \log \det (I + \frac{\mbox{SNR}}{n_t} H
H^{\dagger}) \} ,
\]
which for large SNR has the approximation
\begin{equation} C \ \approx \ \min \{n_t,n_r \}
\log(\mbox{SNR})\label{eq:erg_cap} .
\end{equation}
The space-time code ${\cal X}$ transmits
\[
R \ = \ \frac{1}{T} \log ( \mid {\cal X} \mid )
\]
bits per channel use.   Let $r$ be the normalized rate given by
$R= r \log (\mbox{SNR})$. Following~\cite{ZheTse}, we will refer
to $r$ as the (spatial) multiplexing gain \cite{HeaPau}. From
\eqref{eq:erg_cap}, it is seen that the maximum achievable
multiplexing gain equals $r = \min \{ n_t,n_r\}$. Let the
diversity gain $d(r)$ corresponding to transmission at normalized
rate $r$ be defined by
\[
d(r) \ = \ - \lim_{\text{SNR} \rightarrow \infty }
\frac{\log(P_{e})}{\log (\text{SNR})} ,
\]
where $P_{e}$ denotes the probability of codeword error.  We will
follow the exponential equality notation of \cite{ZheTse} under
which this relationship can equivalently be expressed by \[ P_e \
\dot = \ \text{SNR}^{-d(r)} . \]

A principal result in \cite{ZheTse} is the proof that for a fixed
integer multiplexing gain $r$, and $T\geq n_t+n_r-1$, the maximum
achievable diversity gain $d(r)$ is governed by
\begin{equation}
d(r) \ = \ (n_t-r)(n_r-r) . \label{eq:Zheng-Tse}
\end{equation}
\begin{figure}[ht]
\begin{center}
\centerline{\epsfig{figure=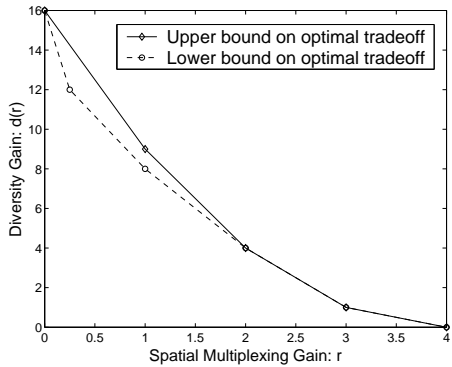,width=60mm}}  \caption{{\it
Upper and Lower Bounds on the D-MG Tradeoff in the case of $4$ transmit and $4$ receive antennas.
\label{fig:d-mg_tradeoff}}
}
\end{center}
\end{figure}
%%%%%%%%%%%%%%%%%%%%%%%%%%%%%%%%%%%%%%%%%%%%%%%%%%%%%%%
The value of $d(r)$ for non-integral values of $r$ is obtained
through straight-line interpolation. For $T< n_t+n_r-1$ only upper
and lower bounds on the maximum possible $d(r)$ are available. The
plots in Fig. \ref{fig:d-mg_tradeoff} are for the case of
$n_r=n_t=T=4$.  The two lines show upper and lower bounds on the
best possible D-MG tradeoff achievable when $T=4$ with any
signalling scheme.

Zheng and Tse\cite{ZheTse} establish that the D-MG tradeoff curve
coincides with the plot of outage probability and that random
Gaussian codes achieve the D-MG tradeoff provided $T \geq
n_t+n_r-1$. They also show that orthogonal space-time block codes
in general, achieve maximum diversity gain but cannot achieve
maximal multiplexing gain.  The reverse is shown to hold in the
case of the vertical Bell-Labs layered space-time architecture
(V-BLAST). Diagonal-BLAST (D-BLAST) in conjunction with MMSE
decoding, is shown to achieve the tradeoff provided one ignores
the overhead due to non-transmission in certain space-time slots.
It is observed in their paper, that apart from the
Alamouti~\cite{Ala} scheme that achieves the D-MG tradeoff in the
case of a single receive antenna, there is no explicitly
constructed coding scheme that achieves the optimum tradeoff for
all $r>0$.

\subsection{Prior work}

In this paper, we shall refer to a ST code that achieves the
upper-bound on the D-MG tradeoff as being a D-MG optimal ST code,
or more simply, an optimal code. The results in \cite{ZheTse}
spurred considerable research activity towards the construction of
optimal codes and some of this is briefly reviewed below.

\subsubsection{Case of Two Transmit Antennas}

For $M$ even, let ${\cal A}_{ \mbox{\tiny QAM}}$ denote the
$M^2$-QAM constellation given by
\begin{equation} \label{A_QAM}
{\cal A}_{\mbox{\tiny QAM}} =  \left\{ a + \imath b \ \mid \
|a|,|b| \leq M-1, \ a,b \ \mbox{odd} \right\}.
\end{equation}

The literature includes several constructions of $(2 \times 2)$ ST
codes that achieve the D-MG tradeoff.  The unnormalized ST code
matrices $X$ in the constructions in
\cite{YaoWor,Yao,DayVar,BelRekVit} share a common structure,
namely
\[
X \ = \ \left[
\begin{array}{cc} x_1 & y_2 \\
y_1 & x_2
\end{array} \right]
\]
where the diagonal and anti-diagonal ``threads''  are unitary
transformations of vectors with QAM components, i.e., where
\[
\left[ \begin{array}{c} x_1 \\ x_2
\end{array} \right] \ = \  S_1 \underline{u}, \ \
\left[
\begin{array}{c} y_1 \\
y_2
\end{array} \right] \ = \ S_2 \underline{v}
\]
with $\underline{u}, \underline{v} \in  {\cal A}_{\tiny QAM}^2$
and where $S_1, S_2$ are unitary. Setting
$M^2=\text{SNR}^{\frac{r}{2}}$ allows the ST code to transmit at
information rate $R= r \log ( \text{SNR})$ bits per channel use.
The constructions differ on the selection of the particular
unitary matrices.

In the Yao-Wornell papers, \cite{YaoWor,Yao} it is shown that by
optimizing amongst the class of unitary matrices corresponding to
rotation of vectors in the complex plane, the minimum determinant
of the difference of two space-time code matrices exceeds
$\frac{1}{2 \sqrt{5}}$ independent of $M$ and hence of the
$\text{SNR}$. This is then used to show that this code achieves
the upper bound on optimum D-MG tradeoff.  This construction is
the first to provide an explicit construction for a code that
achieves the D-MG tradeoff for very value of $n_t,n_r$.

The construction by Dayal and Varanasi, \cite{DayVar} draws on
earlier constructions by Damen et al~\cite{DamTewBel} and El Gamal
and Damen~\cite{GamDam}.  Here the authors optimize the coding
gain through appropriate selection of the matrices $S_1,S_2$. As
with the Yao-Wornell construction, the minimum determinant is
bounded away from zero as $M^2 \rightarrow \infty$. While the D-MG
tradeoff is not explicitly discussed in \cite{DayVar}, by arguing
as in \cite{YaoWor,Yao}, the optimality of the Dayal-Varanasi
construction can be established. The coding gain of the
Dayal-Varanasi construction, especially relevant for lower values
of $\mbox{SNR}$, is shown to improve upon that of the Yao-Wornell
code.

A third construction of a ST code with $n_t=T=2$ that achieves the
D-MG tradeoff is the Golden code construction of Belfiore et al.
\cite{BelRekVit}, so called because of the appearance of the
Golden number $\zeta = \frac{1+\sqrt{5}}{2}$ in the construction.
The Golden code is an example of a class of codes known as perfect
codes~\cite{RekBelVit,OggRekBelVit} and this class of codes is
discussed in greater detail below. Optimality of the Golden code
construction is pointed out in \cite{EliKumPawRajkRajLu}.
Reference~\cite{OggRekBelVit} contains other examples of $( 2
\times 2)$ D-MG optimal codes as well.

A different thread-based construction of D-MG optimal $(2 \times
2)$ code can found in the paper by Liao et al.\cite{LiaWanXia}.

\subsubsection{LAST Codes}

In \cite{GamCaiDam}, El Gamal, Caire and Damen consider a
lattice-based construction of space-time block codes and call
these codes LAST codes.  In this construction, a code matrix $X$
in the space-time code ${\cal C}$ is identified with a $(n_t T
\times 1)$ vector $\underline{x}$ obtained by vertically stacking
the columns of the space-time code matrix.  The construction calls
for a lattice $\Lambda_c$ and a sublattice $\Lambda_s$.  Message
symbols are mapped onto coset representatives $\{ \underline{c}
\}$ of the subgroup $\Lambda_s$ of $\Lambda_c$ that lie within the
fundamental region ${\cal V}_s$ of the sublattice $\Lambda_s$.
Thus the fundamental region of the sublattice serves as a shaping
region for the lattice.  The transmitted vector $\underline{x}$ is
then given by
\[
\underline{x} \ = \ \underline{c} - \underline{u} \pmod{
\Lambda_s} ,
\]
where $\underline{u}$ is a pseudorandom ``dither'' vector chosen
with uniform probability from ${\cal V}_s$. The dither is assumed
to also be known to the receiver.  The lattice pair $\Lambda_s,
\Lambda_c$,  $\Lambda_s \subseteq \Lambda_c$ is drawn from an
ensemble of lattices having good ``covering'' properties, an
example of which can be found in a paper by Loeliger \cite{Loe}.
It is shown that this ensemble of lattices contains a lattice such
that the resultant space-time code, when suitably decoded using
generalized minimum Euclidean distance lattice decoding, achieves
the D-MG tradeoff for all $T \geq n_t+n_r-1$. In actual code
construction, a lattice drawn at random from the ensemble of
lattices is used. A principal advantage of LAST codes is that in
comparison to random Gaussian codes, the decoding is simpler and
does not require searching over the entire codebook.

\subsubsection{Division Algebra- Based Constructions} \label{sec:cda_lit_survey}

\paragraph{ST Codes from Division Algebras}  Space-time code construction from division algebras was
first proposed by Sethuraman and
Rajan~\cite{SetRajITW,SasRajSet_GCOM03,SetRajGC02,SetRajSas} and
independently shortly after, by Belfiore and Rekaya \cite{BelRek}.

In \cite{SetRajSas}, Sethuraman et al. consider the construction
of ST codes from field extensions as well as division algebras
(DA). Two methods of constructing ST codes from DA are presented.
It is shown how Alamouti's code~\cite{Ala} arises as a special
instance of these constructions. A general principle for
constructing CDAs using transcendental elements is given and the
capacity of the space-time block codes (STBCs) obtained via this
construction studied. A second family of CDAs discovered by Brauer
is discussed and applied to construct ST codes.

\paragraph{Non-Vanishing Determinant} The notion of a non-vanishing
determinant (NVD) was introduced by Belfiore and
Rekaya~\cite{BelRek}. The coding gain of a space-time code as
determined by pairwise error probability considerations, is a
function of the determinant of the difference code matrix. It is
therefore of interest to maximize the value of this determinant.
The authors of \cite{BelRek} note that while many constructions of
space-time codes have the property of having a non-zero
determinant, this determinant often vanishes as the SNR increases
and the size of the signal constellation is accordingly increased
to provide increased spectral efficiency. In \cite{BelRek}, the
authors describe an approach for constructing CDA-based square ST
codes whose determinant is bounded from below for all SNR and
hence does not vanish.   NVD code constructions are outlined for
$n_t=T=2^k$ and $n_t=T=3\cdot 2^k$. Example constructions are
provided for $n_t=T=2,3,4$.

\paragraph{Perfect Codes}

In \cite{OggRekBelVit}, Oggier et al. define a square $(n \times
n)$ STBC to be a perfect code
 \footnote{The results of the present paper establish that perfect
codes are D-MG optimal as well.}  if \bit \item the code is a
full-rate, linear-dispersion~\cite{HasHoc} code using $n^2$
information symbols drawn from either a QAM or HEX constellation,
\item the minimum determinant of the code is bounded away from
zero even as $M \rightarrow \infty$, \item the $2M^2$–dimensional
real lattice generated by the vectorized codewords, is either
$\mathbb{Z}^{2·M^2}$ or $A_2^{M^2}$ ($A_2$ is the hexagonal
lattice), and \item each symbol $X_{ij}$ in the code matrix has
the same value of average energy. \eit

Perfect codes have been shown through simulation, to have
excellent performance as judged by codeword error probability. The
authors of \cite{OggRekBelVit} show the existence of perfect
CDA-based space-time codes for dimensions $n=2,3,4,6$. The Golden
code is an example of a perfect code in $2$ dimensions. More
recently, Elia et al. \cite{EliSetKum} show how the perfect code
construction can be generalized to yield perfect codes for any
value of the integer $n$.

\paragraph{Constructions with NVD} In \cite{KirRaj}, square ST codes with the NVD property are
constructed by Kiran and Sundar Rajan for $n_t=T= 2^k, \ 3 \cdot
2^k , \ 2. \cdot 3^k$ or \ $n_t=T=q^k(q - 1)/2$, \ where $q=4s+3$
is a prime. Also contained in this paper, is a lemma that
simplifies the task of identifying the non-norm element $\gamma$
needed in the construction of a CDA having a number field as its
maximal subfield.

\paragraph{Approximate Universality} In \cite{TavVis_CISS,TavVis_ISIT},
Tavildar and Viswanath consider the correlated fading channel
model in which the entries $h_{ij}$  of the matrix $H$ are allowed
to have arbitrary fading distributions.   They show the existence
of permutation codes (codes based on permutations of the QAM
constellation) that achieve the D-MG tradeoff in the case of the
parallel channel (channel where $H$ is diagonal). They note that
by using D-BLAST in conjunction with a permutation code designed
for the parallel channel, one can achieve the D-MG tradeoff of the
general correlated fading channel in the limit as the delay
parameter $T \rightarrow \infty$.  The main result is a sufficient
condition for a ST code to be approximately universal i.e., be
D-MG optimal for every correlated MIMO fading channel. This
sufficient condition is expressed in terms of the product of the
squared-singular values of the difference code matrix.   It is
also shown that in the case of the i.i.d. Rayleigh fading channel,
V-BLAST with a QAM signal constellation achieves the last segment
of the D-MG tradeoff curve while D-BLAST achieves the first
segment when $n_r=2$.

After the initial submission of the present paper, a more detailed
version \cite{TavVis_IT} of \cite{TavVis_ISIT,TavVis_CISS}
containing proofs of all results, has appeared in preprint form.
It is shown in this version (see also \cite{TseVis}) that the
space-time block codes of Yao-Wornell as well as those presented
in the original submission of our present paper here are
approximately universal.

\paragraph{Other Work} Other references relating to the
construction of space-time codes from division algebras include
\cite{CalDigDasAld,KirRaj_Cay,ShaRajSet_CP}. Some analysis of the
D-MG tradeoff of some known constructions can be found in
\cite{KumEliLuPawRajk,SasRajKum,EliKumPawRajkRajLu}. The
rank-distance construction of Lu and Kumar~\cite{LuKum} as well as
subsequent generalizations in \cite{LuKum_Gen_Unified,Ham,Lu} are
all optimal with respect to a different tradeoff known as the
rate-diversity tradeoff which is based on the situation in which
the signal constellation is fixed independent of SNR, and where
pairwise error probability is used as a measure of performance.

\subsection{Principal Results}

A complete solution to the problem of explicitly constructing ST
codes that achieve the D-MG tradeoff of the i.i.d. Rayleigh MIMO
channel is presented in this paper.

The solution is presented in two parts. The first part establishes
the optimality of a class of CDA-based
ST codes having the non-vanishing determinant (NVD) property.  The
codeword matrices in this class of ST code are square i.e., $T =
n_t$ and correspond to minimum-delay ST codes. Prior
constructions~\cite{BelRekVit,KirRaj,RekBelVit} of such ST codes
were restrictive in terms of the values of $n_t$ that could be
accommodated. In the present paper, a general construction for
minimal-delay CDA based ST codes is given for all values of $n_t$.

In the second part, optimal ST constructions are provided for the
rectangular, i.e., $T
> n_t$ case.  Both the square and rectangular constructions
achieve the same tradeoff as do the constructions in the $T \geq
n_t+n_r-1$ case for which the D-MG tradeoff is exactly known from
\cite{ZheTse}. This not only establishes the optimality of the
rectangular constructions, it also extends the range of values of
$T$ for which the D-MG tradeoff is exactly known from $T \geq n_t
+ n_r - 1$ to $T \geq n_t$.

\subsection{Outline}

The optimality of CDA-based ST codes having the NVD property is
established in Section~\ref{sec:CDA_ST}.  This section also
provides some background on division algebras and on ST code
construction from division algebras.  Two constructions of optimal
CDA-based ST codes, valid for all $n_t,n_r$, $T=n_t$ are then
presented in Section~\ref{sec:square_CDA}.
Section~\ref{sec:rectangular} also presents two constructions, but
of optimal rectangular space-time codes valid for all $n_t,n_r$,
$T > n_t$. Appendix I presents a primer on the relevant number
theory.  While the discussion in the paper is focused on
constellations derived from an underlying QAM constellation, the
construction techniques and optimality results also carry over to
the case of the HEX constellation and this is discussed in
Appendix II. Appendix III contains proofs not found elsewhere.

\section{Optimality of CDA-Based ST Codes with NVD Property}
  \label{sec:CDA_ST}

%We begin with some background.

 \subsection{Division Algebras} \label{sec:DA_background} Division algebras are
rings with identity in which every nonzero element has a
multiplicative inverse. As commutative division algebras are
fields, it is the non-commutativity of a division algebra that
serves to differentiate them from fields and as it turns out,
endows certain associated space-time codes with a key non-vanishing
determinant property.

\begin{figure}[h]
\begin{center}
\epsfig{figure=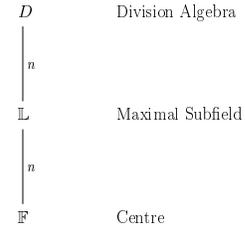,height= 30mm}
\caption{Structure of a Cyclic
Division Algebra\label{fig:tow_CDA}}
\end{center}
\end{figure}

The center $\mathbb{F}$ of any division algebra $D$, i.e., the
subset comprising of all elements in $D$ that commute with every
element of $D$, is a field.  The division algebra is a vector
space over the center \ $\mathbb{F}$ of dimension $n^2$ for some
integer $n$.  A field $\mathbb{L}$ such that $\mathbb{F} \subset
\mathbb{L} \subset D$ and such that no subfield of $D$ contains
$\mathbb{L}$ is called a {\em maximal subfield} of $D$
(Fig.~\ref{fig:tow_CDA}). Every division algebra is also a vector
space over a maximal subfield and the dimension of this vector
space is the same for all maximal subfields and equal to $n$. This
common dimension $n$ is known as the {\em index} of the division
algebra. We will be interested only in the case when the index is
finite.

\begin{ex}[Quaternion Division Algebra] \label{ex:quater}
The classical example of a division algebra is Hamilton's ring
$D=\mathbb{R}(e,i,j,k)$ of quaternions over the real numbers
$\mathbb{R}$, where $e$ is the identity element and $i,j,k$ are
elements satisfying
\begin{eqnarray*}
& i^2=j^2=k^2=-1 & \\
ij=-ji=k,& jk=-kj=i,  &ki=-ik=j.
\end{eqnarray*}
The center of $D$ is the field of real numbers $\mathbb{R}$ and
one maximal subfield is isomorphic to the field of complex numbers
$\mathbb{C}$.  The index $n$ thus equals $2$ in this case.
\end{ex}
\vspace*{0.05in}

\subsection{Cyclic Division Algebras} \label{sec:CDA}

Our interest is in CDA, i.e., division algebras in which the
center $\mathbb{F}$ and a maximum subfield $\mathbb{L}$ are such
that $\mathbb{L}/ \mathbb{F}$ is a cyclic (Galois) extension. CDAs
have a simple characterization that aids in their construction,
see \cite{Alb}, Proposition 11 of \cite{SetRajSas}, or Theorem 1
of \cite{BelRek}.

Let $\mathbb{F}, \ \mathbb{L}$ be number fields, with $\mathbb{L}$
a finite, cyclic Galois extension of $\mathbb{F}$ of degree $n$.
Let $\sigma$ denote the generator of the Galois group
$\text{Gal}(\mathbb{L}/\mathbb{F})$. Let $z$ be an indeterminate
satisfying
\[ \ell z \ = \ z \sigma(\ell) \ \ \ \forall  \ \  \ell \in
\mathbb{L} \ \ \  \mbox{ and } \ \ \  z^n=\gamma, \] for some
non-norm element $\gamma \in \mathbb{F}^*$, by which we mean some
element $\gamma$ having the property that the smallest positive
integer $t$ for which $\gamma^t$ is the relative norm
$N_{\mathbb{L}/\mathbb{F}}(u)$ of some element $u$ in
$\mathbb{L}^*$, is $n$.   Then a CDA
$D(\mathbb{L}/\mathbb{F},\sigma,\gamma) $ with index $n$, center
$\mathbb{F}$ and maximal subfield $\mathbb{L}$ is the set of all
elements of the form
\begin{equation} \label{eq:D_elements}
\sum_{i=0}^{n-1} z^i \ell_i, \ \ \ \ell_i\in \mathbb{L}.
\end{equation}
Moreover it is known that every CDA has this structure. It can be
verified that $D$ is a right vector space (i.e., scalars multiply vectors from the right) over the maximal
subfield $\mathbb{L}$.

\subsection{Space-Time Codes from Cyclic Division Algebras}

A space-time code ${\mathcal{X}}$ can be
associated to $D$ by selecting the set of matrices corresponding
to the matrix representation of elements of a finite subset of
$D$. Note that since these matrices are all square matrices, the
resultant ST code necessarily has $T=n_t$.

The matrix corresponding to an element $d \in D$ corresponds to
the left multiplication by the element $d$ in the division
algebra. Let $\lambda_d$ denote this operation, $\lambda_d: D
\rightarrow D$, defined by
\[ \lambda_d(e) = de , \ \forall \ e \in D. \]
It can be verified that $\lambda_d$ is a ${\mathbb{L}}$-linear
transformation of $D$. From \eqref{eq:D_elements}, a natural
choice of basis for the right-vector space $D$ over $\mathbb{L}$
is $\{ 1,z,z^2,\hdots,z^{n-1} \}$. A typical element in the
division algebra $D$ is $d = \ell_0 + z \ell_1 + \cdots + z^{n-1}
\ell_{n-1}$, where the $\ell_i \in {\mathbb{L}}$. By considering the effect of multiplying
$d \times 1$, $d \times z$, \ldots, $d \times z^{n-1}$, one can show that the ${\mathbb{L}}$-linear transformation
$\lambda_d$ under this basis  has the matrix representation
\begin{equation}
\left[
\begin{array}{ccccc}
  \ell_0 & \gamma \sigma (\ell_{n-1}) & \gamma \sigma^2 (\ell_{n-2}) & \hdots & \gamma \sigma^{n-1} (\ell_1) \\
  \ell_1 & \sigma (\ell_0) & \gamma \sigma^2 (\ell_{n-1}) & \hdots & \gamma \sigma^{n-1} (\ell_2) \\
  \vdots & \vdots & \vdots & \ddots & \vdots \\
  \ell_{n-1} & \sigma (\ell_{n-2}) & \sigma^2 (\ell_{n-3}) & \hdots  & \sigma^{n-1} (\ell_0) \\
\end{array}%
\right] , \label{eq:LeftRegular}
\end{equation}
known as the left regular representation of $d$.

A set of such matrices, obtained by choosing a finite subset of
elements in $D$ constitutes the CDA-based ST code ${\mathcal{X}}$.
The non-commutativity of the CDA endows the codeword matrices with
a key determinant property.

\begin{lem}\label{lem:det_in_F}  Let $A$ denote the $(n \times n)$ matrix
that is the left-regular representation of the element
\[
\psi \ = \ \sum_{i=0}^{n-1} \ell_i z^i, \ \ \ell_i \in \mathbb{L}.
\]
Then $\det (A) \in \mathbb{F}$.
\end{lem}

\begin{proof}
While this result is known (see \cite{Sch} for example), a short
proof is included in Appendix III for the sake of completeness.
\end{proof}

\subsection{Endowing the NVD Property} \label{sec:endowing_nvd}

In this section, we follow ~\cite{BelRek, BelRekVit,OggRekBelVit}
and show how a CDA-based ST code with NVD can be constructed for
the case when the underlying constellation is the QAM
constellation.  The construction can be extended
(see~\cite{EliSetKum}) to other constellations such as the HEX
constellation.  Appendix I provides a primer on the relevant
number theory.  Appendix II provides constructions for the HEX
constellation.

The $\mathcal{A}_{\text{QAM}}$ constellation has the property that
\[ u \in {\cal A}_{\mbox{\tiny QAM}} \ \Rightarrow \ \mid u \mid^2
\leq 2 M^2.
\]
Since
\[
{\cal A}_{\mbox{\tiny QAM}} \ \subseteq \ \mathbb{Q}(\imath) \] it
is natural to consider CDA with center $\mathbb{F} \ = \
\mathbb{Q}(\imath)$.

Let $\mathbb{F} \ = \ \mathbb{Q}(\imath)$, $\mathbb{L}$ be a
$n$-degree cyclic Galois extension $\mathbb{L}/\mathbb{F}$ of
$\mathbb{F}$ and let $\sigma$ be the generator of the Galois group
$\text{Gal}(\mathbb{L}/\mathbb{F})$.   Let ${\cal O}_{\mathbb{F}},
{\cal O}_{\mathbb{L}}$ denote the ring of algebraic integers in
$\mathbb{F}, \mathbb{L}$ respectively. It is known that ${\cal
O}_{\mathbb{F}} = \mathbb{Z}[\imath]$.  Let $\gamma  \ \in \ {\cal
O}_{\mathbb{F}}$, $\gamma \neq 0$, be a non-norm element and
$D(\mathbb{L}/\mathbb{F},\sigma,\gamma) $ denote the associated
CDA.

Let $\{\beta_1, \hdots, \beta_n\}$ form an integral basis for
${\cal O}_{\mathbb{L}}/{\cal O}_{\mathbb{F}}$ and define the set
\[
{\cal A}_{\text{QAM}}(\beta_1, \beta_2,\hdots,\beta_n) \ = \
\left\{ \sum_i a_i \beta_i \mid a_i \in {\cal A}_{\text{QAM}}
\right\} .
\]
Thus ${\cal A}_{\text{QAM}}(\beta_1, \beta_2,\hdots,\beta_n)$ is
the set of all linear combinations of the basis elements $\beta_i$
with coefficients lying in ${\cal A}_{\text{QAM}}$.

Consider the space-time code ${\cal X}$ comprising of matrices corresponding
to the left-regular representation as in \eqref{eq:LeftRegular} of
all elements $d$ in CDA $D$ which are of the form
\[
d = \sum_{i=0}^{n-1} z^i\ell_i\ , \ \ \ \ell_i \in
\mathcal{A}_{\tiny{\text{QAM}}}(\beta_1, \beta_2,\hdots,\beta_n).
\]
From Lemma~\ref{lem:det_in_F}, it follows that the determinant of
every such left-regular representation lies in
$\mathbb{F}=\mathbb{Q}(\imath)$.  But since all entries of the
regular representation lie in the ring ${\cal O}_{\mathbb{L}}$, it
follows that the determinant must moreover, lie in
\[
{\cal O}_{\mathbb{L}} \cap \mathbb{F} \ = \ {\cal O}_{\mathbb{F}}
\ = \ \mathbb{Z[\imath]}. \]  The NVD property of the ST code
constructed now follows since the difference of any two elements
in the CDA is also an element of the CDA and since the magnitude
of any nonzero element in $\mathbb{Z}[\imath]$ is $\geq 1$.

\subsection{Proof of D-MG Optimality of CDA-Based ST Codes having the NVD
Property} \label{sec:proof_of_optimality_given_nvd}

\begin{thm} [Proof of D-MG Optimality] \label{thm:optimal_CDA}
Let $T=n_t=n$. Let the CDA-based ST code ${\cal X}$ be constructed
as above and let ${\cal Z}$ denote the normalized code \[ {\cal Z}
\ = \ \left\{ \theta X \ \mid \ X \in {\cal X} \right\}
\]
where $\theta$ is chosen to ensure that \beq ||\theta X||_F^2 \
\leq \ T \ \text{SNR}, \ \ \text{  all } X \in {\cal X}.
\label{eq:energy_condition} \eeq Then the ST code ${\cal Z}$ is
optimal with respect to the D-MG tradeoff for any number $n_r$ of
receive antennas.
\end{thm}

\begin{proof} We will use the NVD property in conjunction with the sphere bound
to prove optimality of the code. The requirement of transmitting
information at rate $r \log ( \text{SNR} )$ forces $M^2 \ = \
\text{SNR}^{\frac{r}{n}}$. The energy requirement
\eqref{eq:energy_condition} then forces $\theta^2 \ \doteq \
\text{SNR}^{1 - \frac{r}{n}} $.  Let $m=\min \{ n_t,n_r \}$ and
$n_{\Delta}=|n_t-n_r|$.  Let $\theta X_0, \ X_0 \in {\cal X}$ be
the transmitted code matrix and $X_1$ be any other code matrix
distinct from $X_0$. Set $\Delta X = X_0-X_1$. Let \bea \lambda_1
\geq \lambda_2
\geq \cdots \geq \lambda_{n_t} \geq 0  \nonumber \\
0 \leq l_1 \leq l_2 \leq \cdots \leq l_{n_t}
\label{eq:ordering_eigenvalues} \eea denote the eigenvalues of
$H^{\dagger} H$ and $\Delta X \Delta X^{\dagger} $ respectively.
Note that only the first $m$ eigenvalues $\{ \lambda_i \}_{i=1}^m$
are nonzero.  We have the lower bound \bea d_E^2 & = &
\theta^2 \text{Tr} ( H \Delta X \Delta X^{\dagger} H^{\dagger}) \label{eq:opens_to_mismatch_ev_bd}  \\
& \geq  & \theta^2 \sum_{i=1}^{n} l_i \lambda_i \label{eq:mismatch_ev_bd} \\
 & \geq & \theta^2 \sum_{i=1}^k \lambda_i l_i \ \ 1 \leq k \leq m \label{eq:multiple inequalities} \\
& \geq & \theta^2 k \left[ \prod_{i=1}^k \lambda_i
\right]^{\frac{1}{k}} \left[ \prod_{i=1}^k l_i
\right]^{\frac{1}{k}} \ \ \text{ (by AM-GM) }\label{eq:AM-GM_inequality} \\
& \geq & k
\theta^2 \text{SNR}^{- \frac{1}{k} \sum_{i=1}^k \alpha_i} \left[
\frac{\text{SNR}^0}{\text{SNR}^{(n-k)\frac{r}{n}}}
\right]^\frac{1}{k}   \label{eq:using_NVD_trace_prop} \\
%& = & k
% \text{SNR}^{- \frac{1}{k} \sum_{i=1}^k \alpha_i} \left[
%\frac{\theta^{2k}\text{SNR}^0}{\text{SNR}^{(n-k)\frac{r}{n}}}
%\right]^\frac{1}{k}   \label{eq:more_general_result} \\
& \doteq & \text{SNR}^{ -\frac{1}{k} \left\{ \sum_{i=1}^k\alpha_i
+ r-k \right\}} , \ \ 1 \leq k \leq m,  \eea where we have set
$\lambda_i = \text{SNR}^{-\alpha_i}$.

Inequality \eqref{eq:mismatch_ev_bd} is due to K\"{o}se and Wesel
\cite{KosWes} and a proof is provided in Appendix III for the sake
of completeness \footnote{The original submission of this paper
contained an independent proof of this result as we were unaware
at the time of the results in \cite{KosWes}.}.   In
\eqref{eq:using_NVD_trace_prop}, we used the NVD property of the
CDA-based ST code as well as the fact that every eigenvalue $l_i$
is upper bound by $\text{Tr}(\Delta X \Delta X^{\dagger}) \  \dot
\leq  \ M^2 \doteq \text{SNR}^{\frac{r}{n}}$.

By the sphere bound, given the channel matrix $H$, the probability
of codeword error $\text{Pr}({\cal E} | H)$ is upper bounded by
the probability that the additive noise $W$ causes the received
matrix to lie outside a ball of radius $d_{E, \min}/2$ where
$d_{E, \min}$ is the minimum in the set $\{||[\theta
H(X_0-X_1)]||_F \mid X_1 \in {\cal X} \}$.   Our bound for $d_E$
above is independent of the particular pair of codewords
$X_0,X_1$, hence serves as a lower bound to $d_{E, \min}$ and can
be used in place of $d_{E, \min}$ in the sphere bound.

The random variable $||W||_F^2$ is a chi-squared random variable
in $2 n_rT$ dimensions satisfying \beq \text{Pr} \left( ||W||_F^2
> \frac{d_{E, \min}^2}{4} \right) \ = \ \exp  \left( -\frac{d_{E,
\min}^2}{4}\right) \sum_{k=0}^{n_rT-1} \frac{\left( \frac{d_{E,
\min}^2}{4} \right) ^k}{k!} \eeq

Using the density function of the $\{\alpha_i \}$ derived in
\cite{ZheTse}, and averaging over all channel realizations, we
obtain \bean \text{Pr}({\cal E}) & = &
\mathbb{E}_H(\text{Pr}({\cal E}\vert H))
\\
& \leq &  \int_{\underline{\alpha}}  {\cal K}  \ [\log
(\text{SNR})]^m
\exp ( -  (\sum_{i=1}^m \text{SNR}^{-\alpha_i} ) \\
& & \text{SNR}^{-\sum_{i=1}^m \alpha_i [n_{\Delta}+2i-1]} \\
& & \ \ \ \ \ \exp ( - \text{SNR}^{ - \frac{1}{k} (\sum_{i=1}^k
\alpha_i + r-k)} ) \\
& & \left[ \sum_{j=0}^{n_r-1} \frac{\text{SNR}^{-\frac{j}{k}
(\sum_{i=1}^k \alpha_i + r-k)}}{j
!} \right] d \underline{\alpha} \\
& \doteq & \text{SNR}^{ -e(r)} \eean where ${\cal K}$ is a
constant and where
 \bean
& & e(r)  =   \\
& & \inf_{\begin{array}{c}
\alpha_k \geq 0 \\
 \sum_{i=1}^k \alpha_i \geq k-r \\
1 \leq k \leq m \end{array}} \sum_{i=1}^m \alpha_i  (n_{\Delta}+
2(m+1-r)-1) . \eean We find after some work that $e(r)$ is equal
to the piecewise linear function given by $$e(r) =
(n_t-r)(n_r-r)$$ for integral values of $r$.

\end{proof}

A closer examination of the proof of Theorem~\ref{thm:optimal_CDA}
will reveal that the following
more general result is true.  %this determinant condition together

\vspace*{0.2in}

\begin{thm} [Sufficient Condition for Optimality]
\label{thm:suff_condn} Consider a $n_t \times T$ space-time code
$\mathcal{X}$ with $T \geq n_t$ indexed by a rate parameter $r$
such that when the space-time code has size $\text{SNR}^{rT}$ and
when each normalized code matrix $Z \in {\cal Z} \ = \ \{ \theta X
\mid  X \in {\cal X} \}$ satisfies the energy requirement
$||Z||_F^2 \ \leq \ T \ \text{SNR}$, we have
\[\min\limits_{\small
\begin{array}{c} \Delta Z = Z_i-Z_j \neq 0 \\Z_i,Z_j \in {\cal Z}
\end{array}} \det (\Delta Z \Delta Z^{\dag})  \ \dot \geq
\ \text{SNR}^{n_t-r} . \] Then over the quasi-static i.i.d.
Rayleigh-fading channel, the normalized ST code ${\cal Z}$ is
optimal with respect to the D-MG tradeoff for any number $n_r$ of
receive antennas.
\end{thm}
\begin{proof}
Let $\mu_i=\theta^2 l_i$, $\mu_1 \leq \mu_2 \leq \cdots \leq
\mu_n$ denote the eigenvalues of $\Delta Z \Delta Z^{\dagger}$.
From the energy constraint it follows that $\mu_i \ \dot \leq \
\text{SNR}$. Proceeding now as in the proof of
Theorem~\ref{thm:optimal_CDA}, we obtain \bea d_E^2 &  \geq  & k
\left[ \prod_{i=1}^k \lambda_i \right]^{\frac{1}{k}} \left[
\prod_{i=1}^k \mu_i
\right]^{\frac{1}{k}} , \ \ 1 \leq k \leq m  \\
& \geq & k \ \text{SNR}^{- \frac{1}{k} \sum_{i=1}^k \alpha_i}
\left[ \frac{\text{SNR}^{n-r}}{\text{SNR}^{n-k}}
\right]^\frac{1}{k}   \label{eq:using_NVD_trace_prop} \\
%& = & k
% \text{SNR}^{- \frac{1}{k} \sum_{i=1}^k \alpha_i} \left[
%\frac{\theta^{2k}\text{SNR}^0}{\text{SNR}^{(n-k)\frac{r}{n}}}
%\right]^\frac{1}{k}   \label{eq:more_general_result} \\
& \doteq & \text{SNR}^{ -\frac{1}{k} \left\{ \sum_{i=1}^k\alpha_i
+ r-k \right\}} , \ \ 1 \leq k \leq m,  \eea where $m = \min
\{n_r,n_t\}$ as before.  The rest of the proof now proceeds as
before.
\end{proof}

We note that the theorem above can also be derived from the
results in~\cite{TavVis_IT}.

\section{Construction of D-MG Optimal
Square ST Codes Derived from CDA \label{sec:square_CDA}}

By a square ST code, we will mean a ST code in which $T=n_t$ and
we will use $n$ to denote their common value.  In this section, we
will present a construction of D-MG optimal square $(n \times n)$
ST codes derived from CDA. From the discussion in
Section~\ref{sec:CDA_ST}, it follows that the problem of
constructing D-MG optimal ST codes from CDA reduces to one of
identifying cyclic Galois extensions $\mathbb{L}$ of arbitrary
degree $n$ over $\mathbb{F} = \mathbb{Q}(\imath)$ containing a
suitable non-norm element $\gamma$. Prior to the present paper,
these had only been achieved for certain restricted values of $n$.
In \cite{BelRekVit,RekBelVit,OggRekBelVit}, constructions were
provided for values of $n = 2,3,4,6$, while in a more recent
advancement~\cite{KirRaj}, constructions are provided for $n =
2^k, 3.2^k, q^k (q-1)/2$, where $k$ is an arbitrary integer and
$q$ is a prime of the form $4s+3, \ s \in \mathbb{Z}$.

Two general constructions will now be presented, both valid for
all values of the integer $n$.  We begin with a lemma.

\vspace*{0.2in}

\begin{lem} \cite{KirRaj} \label{lem:Kir-BSR}
\ Let $\mathbb{K}$ be a cyclic extension of a number field
$\mathbb{F}$. Let ${\cal O}_{\mathbb{F}}$ denote the ring of
integers of $\mathbb{F}$. Let ${\frak p}$ be a prime ideal of
${\cal O}_{\mathbb{F}}$ that remains inert in the extension
$\mathbb{K}/\mathbb{F}$ and let $\gamma \in {\frak p} \setminus
{\frak p}^2$. Then $\gamma$ is a non-norm element.
\end{lem}

\vspace*{0.2in}

It follows from this lemma, that for constructing D-MG optimal
square $(n \times n)$ ST codes from CDA, it is sufficient to
construct cyclic extensions of $\mathbb{Q}(\imath)$ of degree $n$
such that ${\cal O}_{\mathbb{F}}$ contains a prime ideal ${\frak
p}$ that remains inert in the extension.

\subsection{Construction A}

Let the integer $n$ be factored as follows: \beq n \ = \ 2^{e_0}
\prod_{i=1}^r p_i^{e_i} \ = \ 2^{e_0} n_1 \label{eq:n-fact} \eeq
where the $\{p_i\}$ are distinct odd primes.   Given an integer $m
\geq 3$, we define $\omega_m \ = \ \exp ( \imath \frac{2
\pi}{m})$.

\begin{figure}[!h]
\begin{center}\includegraphics[width=0.8\columnwidth]{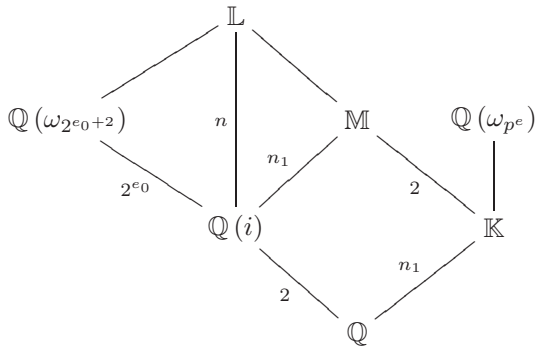}
\caption{The number fields appear in Construction A. $n$ is the
designed cyclic extension.} \label{fig:constn_A}
\end{center}\end{figure}

Construction $A$ begins by identifying two cyclotomic extensions,
one of which contains a cyclic extension $\mathbb{M}$ of
$\mathbb{Q}(\imath)$ of degree $n_1$ and the second of which
contains a cyclic extension $\mathbb{Q}(\omega_{2^{e_0+2}})$ of
$\mathbb{Q}(\imath)$ of degree $2^{e_0}$ (see
Fig.~\ref{fig:constn_A}). Both extensions moreover, contain prime
ideals that remain inert. Then we show how the two extensions can
be made to yield a cyclic extension $\mathbb{L}$ of
$\mathbb{Q}(\imath)$ of degree $n$ also containing an inert prime
ideal.

\begin{prop} \label{prop:Dirichlet} (Dirichlet's theorem) \\ Let $a,m$ be integers such that $1 \leq a
< m$ and $\gcd (a,m) = 1$. Then the arithmetic progression
\[ \{ a, a+m, a+2m, \hdots, a+km, \hdots \} \]
contains infinitely many prime numbers.
\end{prop}

\vspace*{0.2in}

Applying Dirichlet's theorem, Proposition~\ref{prop:Dirichlet}, to
the arithmetic progression
\[ \{ 1, 1+n_1, 1+2n_1, \hdots, 1+kn_1, \hdots \} , \]
we see that we are always guaranteed to find a prime $p$ such that
$n_1 | (p-1)$.  Given this, we define the prime $p$ and the
exponent $e \geq 1$ such that $p^e$ is the smallest prime power
such that $n_1 | (\phi(p^e))$ (where $\phi$ is Euler's totient
function).

\vspace*{0.2in}

\begin{lem} \label{lem:k_is_cyclic} The cyclotomic
extension $\mathbb{Q}(\omega_{p^e})$ contains a subfield
$\mathbb{K}$ that is a cyclic extension of $\mathbb{Q}$ of degree
$n_1$.
\end{lem}
\begin{proof}
Let $G$ denote the Galois group
$\text{Gal}(\mathbb{Q}(\omega_{p^e}) / \mathbb{Q} )$. Then by
Corollary~\ref{cor:cyc_extn_is_cyclic_iff}, $G$ is cyclic of size
$\phi(p^e)=p^{e-1}(p-1)$ and isomorphic to ${\mathbb{Z}}^*_{p^e}$.
Since $n_1 \mid \phi(p^e)$, it follows that $G$ contains a unique
cyclic subgroup $H$ of index $n_1$, i.e., of size $\phi(p^e)/n_1$.

From the fundamental theorem of Galois theory, Proposition
\ref{prop:FTGT}, there is a unique subfield $\mathbb{K}$ of
$\mathbb{Q}(\omega_{p^e})$ which is fixed by the subgroup $H$ such
that ${\mathbb{Q}} \subseteq \mathbb{K} \subseteq
\mathbb{Q}(\omega_{p^e})$. Moreover, $[\mathbb{K}:\mathbb{Q}] \ =
\ n_1$ and $\text{Gal}(\mathbb{K}/\mathbb{Q}) \thickapprox G/H$.
Thus to show that $\mathbb{K}/\mathbb{Q}$ is cyclic, it is
sufficient to show that $G/H$ is cyclic and this follows from
Lemma~\ref{lem:quotient_gp_is_cyclic}.
\end{proof}

%\vspace*{0.2in}
%
%\begin{figure}[!h]
%\xymatrix{ & {\mathbb L} \ar@{-}[ld] \ar@{-}[dd]_{n}\ar@{-}[rd] & & \\
%\Q \left( \omega_{2^{e_0+2}} \right) \ar@{-}[rd]_{2^{e_0}}& &
%{\mathbb M}
%\ar@{-}[rd]_{2} \ar@{-}[ld]_{n_1} &  \Q \left( \omega_{p^{e}} \right) \ar@{-}[d]\\
%& \Q \left( i \right) \ar@{-}[rd]_{2} & & {\mathbb K} \ar@{-}[ld]_{n_1}  \\
%& & \Q & } \caption{The number fields appear in Construction A.
%$n$ is the designed cyclic extension.}
%\label{fig:constn_A}
%\end{figure}
%
%\vspace*{0.2in}

\vspace*{0.2in}

\vspace*{0.2in}

\begin{lem} \label{lem:inert_in_odd_case} There exists a rational prime
$q_1$ such that the ideal $(q_1)$ is inert in
$\mathbb{Q}(\omega_{p^e})/\mathbb{Q}$.
\end{lem}
\begin{proof} Let $\rho$ be a generator of the cyclic group $\mathbb{Z}_{p^e}^{*}$.
Hence $\rho$ has order $\phi( p^e)$.  By Dirichlet's theorem,
Proposition~\ref{prop:Dirichlet}, applied to the arithmetic
progression
\[ \{ \rho, \rho+p^e, \rho+2p^e, \hdots, \rho + k p^e, \hdots \} \]
there exists a prime $q_1 \equiv \rho \pmod{p^e}$. From
Lemma~\ref{lem:decomp_prime_power_cyc_fields} it follows that
$(q_1)$ is inert in
$\mathbb{Q}(\omega_{p^e})/\mathbb{Q}$.
\end{proof}

\vspace*{0.2in}

Our next aim is to construct a cyclic extension of
$\mathbb{Q}(\imath)$ of degree $2^{e_0}$ that contains an inert prime ideal.   If $e_0=0$, then
$\mathbb{Q}(\imath)$ itself is the desired extension and we are done.

\begin{figure}[!h]
\includegraphics[width=0.9\columnwidth]{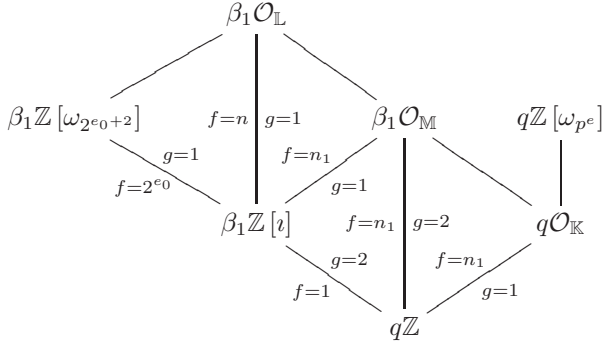} \caption{Prime ideal decomposition in the number
fields of Construction A. Note that although $q\mathbb{Z}$
decomposes into the product of two prime ideals in the extensions
$\mathbb{Q}(\imath)/\mathbb{Q}$ and $\mathbb{M}/\mathbb{Q}$, we
have shown only one of the two prime ideals in the figure for the
sake of clarity, namely the one generated by $\beta_1$.
 \label{fig:ideal_decomp}}
\end{figure}

For $e_0 \geq 1$, the group $\mathbb{Z}^*_{2^{e_0+2}}$ is not
cyclic by Lemma~\ref{lem:zmstar_is_cyclic} and so one cannot hope
to find a rational prime $q$ which remains inert in the extension
$\mathbb{Q}(\omega_{2^{e_0+2}})/\mathbb{Q}$.   However, it turns
out that one can always find a prime $q$ such that the ideal $(q)$
splits into the product $(q)=\beta_1 \beta_2$ of prime ideals
$\beta_i, i=1,2$ in the extension $\mathbb{Q}(\imath)/\mathbb{Q}$
with each of the prime ideals $\beta_i$ remaining inert in the
extension $\mathbb{Q}(\omega_{2^{e_0+2}})/\mathbb{Q}(\imath)$ and
this turns out to be sufficient for our purposes (see
Fig.~\ref{fig:ideal_decomp}). We begin by showing  that the
extension $\mathbb{Q}(\omega_{2^{e_0+2}})/\mathbb{Q}(\imath)$ is
cyclic.

\vspace*{0.2in}

\begin{lem} \label{lem:even_is_cyclic} ${\mathbb{Q}}(\omega_{2^{e_0+2}})/{\mathbb{Q}}(\imath)$
is a cyclic Galois extension of degree $2^{e_0}$.
\end{lem}
\begin{proof} From Lemma~\ref{lem:zevenstar_is_not_cyclic},
the maximal order of an element in ${\mathbb{Z}}_{2^{e_0+2}}^{*}$ equals $2^{e_0}$.
The element $5$ in ${\mathbb{Z}}_{2^{e_0+2}}^{*}$
has this maximal possible order.  This follows since
every element must have order dividing $\phi(2^{e_0+2})=2^{e_0+1}$ and
$5^{2^{e_0-1}} \equiv (1+2^2)^{2^{e_0-1}} \equiv 1+2^{e_0-1}.2^2
\not\equiv 1 \pmod{2^{e_0+2}}$.  Next, consider the automorphisms
$\sigma_k, 0 \leq k \leq 2^{e_0}-1$, of ${\mathbb
Q}(\omega_{2^{e_0+2}})/{\mathbb Q}$ given by
\[
\sigma_k(\omega_{2^{e_0+2}}) \ = \ \omega_{2^{e_0+2}}^{5^k}, \ \ 0
\leq k \leq 2^{e_0}-1.
\]
These automorphisms form a cyclic group of order $2^{e_0}$ and the
fixed field of this group is ${\mathbb{Q}}(\imath)$ since
$\imath^{5^k}=\imath$, all $k$. It follows that
${\mathbb{Q}}(\omega_{2^{e_0+2}})/ {\mathbb{Q}}(\imath)$ is cyclic
of degree $2^{e_0}$.
\end{proof}

\vspace*{0.2in}

\begin{lem} \label{lem:inert_in_even_case} There exists a prime ideal
$\beta$ such that $\beta$ is inert in
$\mathbb{Q}(\omega_{2^{e_0+2}})/\mathbb{Q}(\imath)$.
\end{lem}
\begin{proof} Let $p$ be a prime such that $p \equiv 1 \pmod{4}$.
Then it follows from Lemma~\ref{lem:ram_index_gt_1} and
Lemma~\ref{lem:decomp_prime_power_cyc_fields} that both the
ramification index $e$ as well as the relative degree $f$ of the
prime ideal $p \mathbb{Z}[\imath]$ in the extension
$\mathbb{Q}(\imath) / \mathbb{Q}$ equal $1$.  It follows that the
index $g$ of the decomposition group of $p \mathbb{Z}$ equals $2$,
i.e., that the ideal $p \mathbb{Z}[\imath]$ splits into the
product
\[
p \mathbb{Z}[\imath] \ = \ \beta_1 \beta_2
\]
of distinct prime ideals $\beta_1, \beta_2$.

Next let us impose the further condition that $ p \equiv 5
\pmod{2^{e_0+2}}$ which is consistent with $p \equiv 1 \pmod{4}$.
Consider the decomposition of $p\mathbb{Z}$ in the larger
extension $\mathbb{Q}(\omega_{2^{e_0+2}}) / \mathbb{Q}$. From the
proof of Lemma~\ref{lem:even_is_cyclic}, it follows that the order
of $p \pmod{2^{e_0+2}}$ equals $2^{e_0}$. Thus the relative degree
$f$ of the prime ideal $p \mathbb{Z}[\omega_{2^{e_0+2}}]$ equals
$2^{e_0}$.  Since $\phi(2^{e_0+2})=2^{e_0+1}$ and the ideal $p
\mathbb{Z}[\omega_{2^{e_0+2}}]$ is unramified, it follows from
equation \eqref{eq:efg_equals_n} that the index $g$ of the
corresponding decomposition group equals $2$ so that the ideal
generated by $p$ in the ring of integers
$\mathbb{Z}[\omega_{2^{e_0+2}}]$ of
$\mathbb{Q}(\omega_{2^{e_0+2}})$, factors into the product of two
prime ideals.

It follows now from \eqref{eq:f_eq_f_1_times_f_2} that both prime
ideals $\beta_i$ remain inert in the extension
$\mathbb{Q}(\omega_{2^{e_0+2}})/\mathbb{Q}(\imath)$.
\end{proof}

%\begin{figure}[!h]
%\xymatrix{ & \beta_1 {\cal O}_{\mathbb L} \ar@{-}[ld] \ar@{-}[dd]_{f=n}^{g=1}\ar@{-}[rd] & & \\
%\beta_1 \Z \left[ \omega_{2^{e_0+2}} \right]
%\ar@{-}[rd]_{f=2^{e_0}}^{g=1}& & \beta_1 {\cal O}_{\mathbb M}
%\ar@{-}[rd] \ar@{-}[ld]_{f=n_1}^{g=1} \ar@{-}[dd]_{f=n_1}^{g=2}&
%q \Z \left[ \omega_{p^e} \right] \ar@{-}[d]\\
%& \beta_1 \Z  \left[ \imath \right] \ar@{-}[rd]_{f=1}^{g=2} &
%& q {\cal O}_{\mathbb K} \ar@{-}[ld]_{f=n_1}^{g=1}  \\
%& & q \Z & } \caption{Prime ideal decomposition in the number
%fields of Construction A.
%Note that although $q\mathbb{Z}$ decomposes into the product of two prime ideals in the extensions $\mathbb{Q}(\imath)/\mathbb{Q}$ and $\mathbb{M}/\mathbb{Q}$, we have shown only one of the two prime ideals in the figure for the sake of clarity, namely the one generated by $\beta_1$.
% \label{fig:ideal_decomp}}
%\end{figure}

%\begin{figure}[!t]
%\begin{center}
%\includegraphics[width=90mm]{Ideal_decomposition.eps}
%\caption{Prime ideal decomposition in the number fields of Construction A.
%Note that although $q\mathbb{Z}$ decomposes into the product of two prime ideals in the extensions $\mathbb{Q}(\imath)/\mathbb{Q}$ and $\mathbb{M}/\mathbb{Q}$, we have shown only one of the two prime ideals in the figure for the sake of clarity, namely the one generated by $\beta_1$.
% \label{fig:ideal_decomp}}
%\end{center}
%\end{figure}
%

\vspace*{0.2in}

\begin{thm}  [Construction A] \label{thm:constn_A}  Let
\beq n \ = \  2^{e_0} n_1
 \eeq where $n_1$ is odd.  Let $p^e$ be the smallest prime
 power such that $n_1 | \phi(p^e)$.  Let $G$ be the Galois group of
 $\mathbb{Q}(\omega_{p^e}) / \mathbb{Q}$.  Let $H$ be a subgroup of $G$
 of size $\phi(p^e)/n_1$.  Let $\mathbb{K}$ be the fixed field of $H$.
 Let $\mathbb{M}$ be the compositum of $\mathbb{K}$ and $\mathbb{Q}(\imath)$ and
 $\mathbb{L}$ the compositum of $\mathbb{M}$ and $\mathbb{Q}(\omega_{2^{e_0+2}})$.
 Then $\mathbb{L}$ is the desired cyclic extension of $\mathbb{Q}(\imath)$ of degree $n$ (Fig.~\ref{fig:constn_A}).

Let $\rho \in \mathbb{Z}^*_{p^e}$ be a generator of the cyclic
group $\mathbb{Z}_{p^e}^{*}$.  Let $q$ be a rational prime such
that \bea
q & = & \left\{ \begin{array}{lrl} \rho & \pmod{p^e} \\
 \\ 5 & \pmod{2^{e_0+2}} \end{array} \right. .
\label{eq:desired_prime_q}  \eea Let $\beta$ be a
prime ideal of $\mathbb{Z}[\imath]$ lying above $q\mathbb{Z} $ in
$\mathbb{Q}(\imath) / \mathbb{Q}$. Then $\beta$ is the desired prime that remains inert in the
extension $\mathbb{L} / \mathbb{Q}(\imath)$.
\end{thm}

\begin{proof}  From Lemma~\ref{lem:k_is_cyclic} we know that $\mathbb{K}/\mathbb{Q}$ is a
cyclic extension of odd degree $n_1$.  It is clear that
$\mathbb{Q}(\imath)/\mathbb{Q}$ is cyclic of degree $2$.  It
follows from Lemma~\ref{lem:compositum_is_cyclic} that the
compositum $\mathbb{M}=\mathbb{K} \cdot \mathbb{Q}(\imath)$ is a cyclic
extension of $\mathbb{Q}$ of degree $2 n_1$.  Since $\mathbb{M}/
\mathbb{Q}$ is cyclic, it follows that the extension $\mathbb{M} /
\mathbb{Q}(\imath)$ is cyclic of degree $n_1$ as well. From
Lemma~\ref{lem:even_is_cyclic} we know that
$\mathbb{Q}(\omega_{2^{e_0+2}}) / \mathbb{Q}(\imath)$ is cyclic of degree $2^{e_0}$. It
follows now from a second application of
Lemma~\ref{lem:compositum_is_cyclic} that $\mathbb{L} /
\mathbb{Q}(\imath)$ is cyclic of degree $n$ as desired.

To prove that the ideal $\beta$ is inert in the extension
$\mathbb{L} / \mathbb{Q}(\imath)$, we note first of all that by
the Chinese Remainder theorem and Dirichlet's theorem,
Proposition~\ref{prop:Dirichlet}, a prime $q$ satisfying
\eqref{eq:desired_prime_q} is guaranteed to exist. Next we observe
that \ben \item  the index of the decomposition group of
$q\mathbb{Z}$ in $\mathbb{Q}(\imath)/\mathbb{Q}$ equals $2$. Hence
from \eqref{eq:g_eq_g_1_times_g_2}, the index of the decomposition
group of $q\mathbb{Z}$ in $\mathbb{M}/\mathbb{Q}$ must be a
multiple of $2$, \item the relative degree of $q\mathbb{Z}$ in the
extension $\mathbb{K}/\mathbb{Q}$ equals $n_1$, hence the relative
degree of $q\mathbb{Z}$ in the extension $\mathbb{M}/\mathbb{Q}$
must be a multiple of $n_1$.  \een On the other hand, we have that
the degree of the extension $\mathbb{M}/\mathbb{Q}$ equals $2n_1$
and hence it follows from equation \eqref{eq:efg_equals_n} of
Appendix I that the index of the decomposition group of
$q\mathbb{Z}$ in $\mathbb{M}/\mathbb{Q}$ equals $2$ and the
relative degree of $q\mathbb{Z}$ in the extension
$\mathbb{M}/\mathbb{Q}$ equals $n_1$.

Since the index of the decomposition group of $q \mathbb{Z}$ in
$\mathbb{Q}(\imath)/\mathbb{Q}$ equals $2$, $q \mathbb{Z}$ factors
into the product of two ideals $\beta_1 \beta_2$ in this
extension.  Without loss of generality, we set $\beta=\beta_1$. An
application of equation \eqref{eq:f_eq_f_1_times_f_2} of Appendix
I, then tells us that the relative degree of the ideal $\beta$ in
the extension $\mathbb{M}/\mathbb{Q}(\imath)$ equals $n_1$, i.e.,
that the ideal is inert in this extension.  Since the ideal
$\beta$ is inert even in the extension
$\mathbb{Q}(\omega_{2^{e_0+2}})/\mathbb{Q}(\imath)$, it follows
once again from \eqref{eq:f_eq_f_1_times_f_2}, that the ideal
$\beta \mathbb{Z}[\imath]$ remains inert in the extension
$\mathbb{L}/\mathbb{Q}(\imath)$ where $\mathbb{L}$ is the
compositum of the fields $\mathbb{Q}(\omega_{2^{e_0+2}})$ and
$\mathbb{M}$ (see Fig.~\ref{fig:ideal_decomp}).
\end{proof}

\subsection{Construction B}

\begin{figure}
\[\includegraphics[width=0.9\columnwidth]{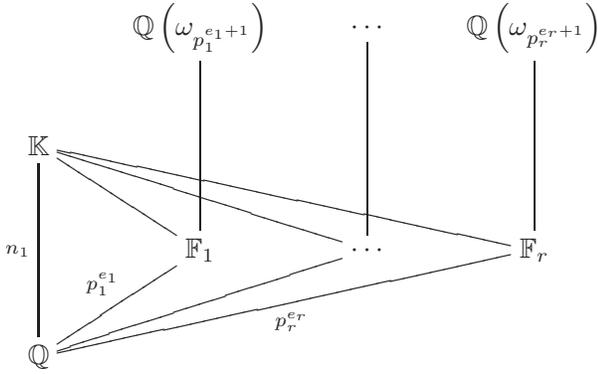}\]
  \caption{Constructing the cyclic extension $\mathbb{K}/\mathbb{Q}$ of Construction B.  \label{fig:constn_B_K_part}}
\end{figure}

\begin{figure}[!h]
\[\includegraphics[width=0.7\columnwidth]{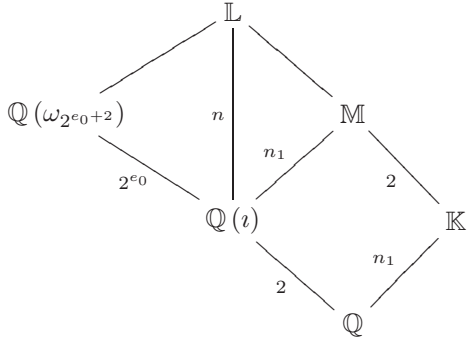}\] \caption{Construction B with $\mathbb{K}$ constructed
as in Fig.~\ref{fig:constn_B_K_part}.}
\label{fig:constn_B_rest_of_it}
\end{figure}

\begin{thm} [Construction B] \label{thm:constn_B}  Let
\beq n \ = \ 2^{e_0} \prod_{i=1}^r p_i^{e_i} \ = \ 2^{e_0} n_1
 \eeq where $p_i$ are distinct odd primes.
 Let $\rho_i, 1 \leq i \leq r $ be elements of $\mathbb{Z}_{p_i^{e_i+1}}^{*}$ having
 maximum possible multiplicative order $p_i^{e_i}$.
 Let $\rho$ be the element of $\mathbb{Z}_n^{*}$ satisfying
 \bea
 \rho & \equiv & \left\{ \begin{array}{l} \rho_i \pmod{p_i^{e_i+1}}, \  \ 1 \leq i \leq r \\
  5 \pmod{2^{e_0+2}} . \end{array} \right.
 \eea
Let $H_i$ be the unique cyclic subgroup of
$\text{Gal}\left(\mathbb{Q}(\omega_{p_i^{e_i+1}} )/ \mathbb{Q}
\right)$ of index $p_i^{e_i}, \ 1 \leq i \leq r$.    Let
$\mathbb{F}_i, 1 \leq i \leq r$ be the fixed field of $H_i$.
 Let $\mathbb{K}$ be the compositum of the $(r+1)$ fields $\mathbb{Q}(\imath)$,
 $\{ \mathbb{F}_i \}_{i=1}^r$ and let $\mathbb{L}$
 be the compositum of $\mathbb{K}$ and
$\mathbb{Q}(\omega_{2^{e_0+2}})$
(Fig.~\ref{fig:constn_B_K_part},~\ref{fig:constn_B_rest_of_it}).
 Then
 \ben
\item there exists a prime $q \ \equiv \ \rho \pmod{n}$
 \item
$\mathbb{L}$ is a cyclic extension of $\mathbb{Q}(\imath)$ of
degree
 $n$
 \item the ideal generated by $q\mathbb{Z}$ factors into the
 product of two prime distinct ideals $\beta_i, i=1,2$ in the ring of algebraic
 integers of $\mathbb{Z}[\imath]$, i.e.,
 \[
 q\mathbb{Z} \ = \ \beta_1 \beta_2
 \]
 \item each of the two ideals $\beta_i$ remains inert in the field
 extension $\mathbb{L} / \mathbb{Q}(\imath)$
 \een
 \end{thm}

\begin{proof}
The same argument used in the proof of Lemma~\ref{lem:k_is_cyclic}
to show that the extension $\mathbb{K} / \mathbb{Q}$ is cyclic,
shows that each of the fields $\mathbb{F}_i$ are cyclic extensions
of $\mathbb{Q}$ of degree $p_i^{e_i}, 1 \leq i \leq r$ and that $q
\mathbb{Z}$ remains inert in the extension
$\mathbb{F}_i/\mathbb{Q}$.  Since the extensions
$\mathbb{F}_i/\mathbb{Q}$ have degrees that are relatively prime,
it follows that $\mathbb{K}/\mathbb{Q}$ is a cyclic extension of
degree $n_1=\prod_{i=1}^r p_i^{e_i}$.  Moreover, since $q
\mathbb{Z}$ remains inert in each of the extensions
$\mathbb{F}_i/\mathbb{Q}$, it follows from
\eqref{eq:f_eq_f_1_times_f_2} and \eqref{eq:efg_equals_n} that $q
\mathbb{Z}$ remains inert in the extension
$\mathbb{K}/\mathbb{Q}$.

We are now in a similar situation as in the case of Construction
A. The only difference is the manner in which the field
$\mathbb{K}$ was constructed.  Thus by following the remainder of
the proof of Theorem~\ref{thm:constn_A}, the proof of the present
proposition follows.
\end{proof}

%\begin{figure*}
%\[
%\xymatrix{ & {\mathbb L} \ar@{-}[ld]_{n_1} \ar@{-}[rd]^{2^{e_0+1}}
%\ar@{-}[dd]_{n} & & \Q \left( \omega_{p_1^{e_1+1}}
%\right)\ar@{-}[dd]& \cdots \ar@{-}[dd]
%& \Q \left( \omega_{p_r^{e_r+1}}\right)\ar@{-}[dd] \\
%\Q \left( \omega_{2^{e_0+2}} \right) \ar@{-}[rd]_{2^{e_0}} &&
% \ar@{-}[ld] {\mathbb K} \ar@{-}[dd]_{n_1} \ar@{-}[rd] \ar@{-}[rrd] \ar@{-}[rrrd] & &&\\
%&\Q (i) \ar@{-}[rd]_{2} && \mathbb{F}_1 \ar@{-}[ld]_{p_1^{e_1}} &
%\cdots \ar@{-}[lld]& \mathbb{F}_r\ar@{-}[llld]^{p_r^{e_r}}\\
% &&\Q &&&}
%\]
%  \caption{The number fields appearing in
%Construction B.  \label{fig:constn_B}}
%\end{figure*}

%\begin{figure}
%\[
%\xymatrix{  & & \Q \left( \omega_{p_1^{e_1+1}} \right)\ar@{-}[dd]&
%\cdots \ar@{-}[dd]
%& \Q \left( \omega_{p_r^{e_r+1}}\right)\ar@{-}[dd] \\
% &
% {\mathbb K} \ar@{-}[dd]_{n_1} \ar@{-}[rd] \ar@{-}[rrd] \ar@{-}[rrrd] & &&\\
% && \mathbb{F}_1 \ar@{-}[ld]_{p_1^{e_1}} &
%\cdots \ar@{-}[lld]& \mathbb{F}_r\ar@{-}[llld]^{p_r^{e_r}}\\
% &\Q &&&}
%\]
%  \caption{Constructing the cyclic extension $\mathbb{K}/\mathbb{Q}$ of Construction B.  \label{fig:constn_B_K_part}}
%\end{figure}

%\begin{figure}[!h]
%\xymatrix{ & {\mathbb L} \ar@{-}[ld] \ar@{-}[dd]_{n}\ar@{-}[rd] & & \\
%\Q \left( \omega_{2^{e_0+2}} \right) \ar@{-}[rd]_{2^{e_0}}& &
%{\mathbb M}
%\ar@{-}[rd]_{2} \ar@{-}[ld]_{n_1} &   \\
%& \Q \left( \imath \right) \ar@{-}[rd]_{2} & & {\mathbb K} \ar@{-}[ld]_{n_1}  \\
%& & \Q & } \caption{Construction B with $\mathbb{K}$ constructed
%as in Fig.~\ref{fig:constn_B_K_part}.}
%\label{fig:constn_B_rest_of_it}
%\end{figure}

%\begin{figure*}
%\[
%\xymatrix{ & {\mathbb L} \ar@{-}[ld]_{n_1} \ar@{-}[rd]^{2^{e_0+1}}
%\ar@{-}[dd]_{n} & & \Q \left( \omega_{p_1^{e_1+1}}
%\right)\ar@{-}[dd]& \cdots \ar@{-}[dd]
%& \Q \left( \omega_{p_r^{e_r+1}}\right)\ar@{-}[dd] \\
%\Q \left( \omega_{2^{e_0+2}} \right) \ar@{-}[rd]_{2^{e_0}} &&
% \ar@{-}[ld] {\mathbb K} \ar@{-}[dd]_{n_1} \ar@{-}[rd] \ar@{-}[rrd] \ar@{-}[rrrd] & &&\\
%&\Q (i) \ar@{-}[rd]_{2} && \mathbb{F}_1 \ar@{-}[ld]_{p_1^{e_1}} &
%\cdots \ar@{-}[lld]& \mathbb{F}_r\ar@{-}[llld]^{p_r^{e_r}}\\
%& & & & \\
% &&\Q &&&}
%\]
%  \caption{The number fields appearing in
%Construction B.  \label{fig:constn_B}}
%\end{figure*}

\begin{table}[h] \caption{Non-Norm Elements for Use in
Construction A}
\begin{center}
\begin{tabular}{|c|c|c|c|}
  \hline
  No. $n_t$ of Antennas & $p^e$ &  $q$ & Non-norm `$\gamma$' \\
  \hline
  2 & 8& 5 & $(2+\imath)$ \\
  \hline
  3 & 7 & 5 & $(2+\imath)$ \\
  \hline
  4 & 8& 5 & $(2+\imath)$ \\
    \hline
  5 & 11& 13 & $(3+2\imath)$ \\
    \hline
  6 & 7& 5 & $(2+\imath)$ \\
    \hline
  7 & 29& 37 & $(6+ \imath)$ \\
    \hline
  8 & 8& 5 & $(2+\imath)$ \\
    \hline
  9 & 19 & 29 & $(5+2\imath)$ \\
    \hline
  10 & 11& 13 & $(3+2\imath)$ \\
    \hline
  11 & 23 & 5 & $(2+\imath)$ \\     \hline
  12 & 7 & 5 & $(2+ \imath)$ \\     \hline
13 & 53 & 5 & $(2 +  \imath)$ \\     \hline 14 & 29 & 36 & $(6 + \imath)$ \\
\hline 15 & 31 & 53 & $(7+2\imath)$ \\     \hline
   16 & 8 & 5 & $(2 +\imath)$ \\     \hline
  17 & 103 & 5 & $(2+\imath)$ \\     \hline
   18 & 19 & 13 & $(3+2 \imath)$ \\     \hline
19 & 191 & 29 & $(5+2\imath)$ \\     \hline
  20 & 11 & 13 & $(3 + 2 \imath)$ \\
  \hline
\end{tabular}
\end{center}
\label{tab:gamma_table_A}
\end{table}

\vspace*{0.2in}

\begin{table}[h]
\caption{Non-Norm Elements for Use in Construction B}
\begin{center}
\begin{tabular}{|c|c|c|}
  \hline
  No. $n_t$ of Antennas & prime `$q$' & Non-norm `$\gamma$' \\
  \hline
  2 & 5 & $(2+\imath)$ \\
  \hline
  3 & 5 & $(2+\imath)$ \\
  \hline
  4 & 5 & $(2+\imath)$ \\
    \hline
  5 & 13 & $(3+2\imath)$ \\
    \hline
  6 & 5 & $(2+\imath)$ \\
    \hline
  7 & 5 & $(2+\imath)$ \\
    \hline
  8 & 5 & $(2+\imath)$ \\
    \hline
  9 & 5 & $(2+\imath)$ \\
    \hline
  10 & 13 & $(3+2\imath)$ \\
    \hline
  11 & 13 & $(3+2\imath)$ \\     \hline
  12 & 5 & $(2 + \imath)$ \\     \hline
13 & 37 & $(6 + \imath)$ \\     \hline
14 & 5 & $(2 + \imath)$ \\
\hline 15 & 113 & $(7 + 8\imath)$ \\     \hline
  16 & 5 & $(2 + \imath)$ \\     \hline
  17 & 5 & $(2 + \imath)$ \\     \hline
   18 & 5 & $(2 + \imath)$ \\     \hline
19 & 13 & $(3 + 2\imath)$ \\     \hline
  20 & 37 & $(6 + \imath)$ \\
  \hline
\end{tabular}
\end{center}
\label{tab:gamma_table_B}
\end{table}
Having explicitly described the method for constructing $\gamma$
for any number of antennas, we proceed in Tables
\ref{tab:gamma_table_A}, \ref{tab:gamma_table_B}, to present
example values of $\gamma$ for $n_t$ in the range $2 \leq n_t \leq
20$ corresponding to Constructions A and B respectively.

\section{Constructions For The Rectangular Case} \label{sec:rectangular}

For the purposes of simplifying the exposition in this section, we
will use the term {\em clearly optimal} $(n_t \times T)$ ST code
to refer to a normalized $n_t \times T$ space-time code
$\mathcal{Z}$ with $T \geq n_t$, indexed by a rate parameter $r$
such that when \bit \item[(i)] the space-time code has size
$\text{SNR}^{rT}$ and when \item[(ii)] each normalized code matrix
$Z \in {\cal Z}$ satisfies $$||Z||_F^2 \ \leq \ T \ \text{SNR},$$
\item[(iii)] we have
\[\min\limits_{\small
\begin{array}{c} \Delta Z = Z_i-Z_j \neq 0 \\Z_i,Z_j \in {\cal Z}
\end{array}} \det (\Delta Z \Delta Z^{\dag})  \ \dot \geq
\ \text{SNR}^{n_t-r} . \] \eit It follows from
Theorem~\ref{thm:suff_condn}, that a clearly optimal ST code is
D-MG optimal for any number $n_r$ of receive antennas.

Two general techniques will now be presented that enable the
construction of clearly optimal rectangular $(n_t \times T)$ ST
codes for every pair $(n_t,T)$ with $T
> n_t$ from clearly optimal square ST codes of the appropriate dimension.

Both techniques will be shown to yield ST codes that achieve the
upper bound on the optimal D-MG tradeoff for all $T\geq n_t$ (and
all $n_r$ for a given $T$) given by \beq d(r) \ = \ (n_t-r)(n_r-r)
\label{eq:D-MG_Tradeoff_rect} \eeq for integer $r$ and by
straight-line interpolation for non-integer values. This will not
only establish the D-MG optimality of these construction
techniques, but show in addition that for all $T \geq n_t$, the
D-MG tradeoff is given precisely by \eqref{eq:D-MG_Tradeoff_rect}.
Previously, from \cite{ZheTse}, this was known to hold only for
all $T \geq n_t+n_r-1$.

The first technique, which we term the row-deletion construction,
shows that a clearly optimal square ST code remains clearly
optimal even if an arbitrary number of rows from the matrix is
deleted.

The second construction, called the Cartesian-product
construction, shows that codewords from a collection of clearly
optimal $(n_t \times T_i)$ ST codes can be horizontally stacked to
yield a clearly optimal ST code of larger length $T=\sum_i T_i$.

\subsection{The Row-Deletion Construction} \label{sec:rect_row_deletion}

\begin{thm} [Row-Deletion Construction] \label{thm:row_deletion}
Let $T > n_t$.  Let ${\mathcal{Z}}_S$ be a clearly optimal $(T
\times T)$ square ST code.
 Next, let ${\cal Z}_R$ be the $(n_t \times T)$ rectangular ST code obtained
by deleting a particular set of $(T-n_t)$ rows from every code
matrix $Z \in {\cal Z}_S$.   Then the $( n_t \times T)$ ST code
${\cal Z}_R$ is also clearly optimal.
\end{thm}

\begin{proof}
We note first that the ST code ${\cal Z}_S$ has the property that
the message symbols can be uniquely recovered given either a
single row or a single column of the code matrix.   This follows
since no two code matrices in ${\cal Z}_S$ can agree in any single
row or column, for, otherwise, their difference would have
determinant equal to zero.

%For example, from the row
%\begin{equation}
%\theta\left[
%\begin{array}{cccc}
%  \ell_1 & \sigma (\ell_0) &  \hdots & \gamma \sigma^{n-1} (\ell_2) \\
%\end{array}%
%\right] ,
%\end{equation}
%one can uniquely determine the information-bearing symbols
%$\{\ell_i\}$.
Next we observe that if $Z_R$ is a $(n_t \times T)$ matrix
obtained from a $(T \times T)$ ST code matrix $Z_S \in {\cal
Z}_S$, by deleting some $(T-n_t)$ rows, then $\Delta Z_R \Delta
Z_R^{\dagger}$ is a $(n_t \times n_t)$ principal submatrix of
$\Delta Z_S \Delta Z_S^{\dagger}$.  Let \bea & & \mu_1 \leq \mu_2
\leq
\cdots \leq \mu_{n_t} \\
& & \nu_1 \leq \nu_2 \leq \cdots \leq \nu_{n_t} \leq \cdots \leq
\nu_T \eea be the ordered eigenvalues of $\Delta Z_R \Delta
Z_R^{\dagger}$ and $\Delta Z_S \Delta Z_S^{\dagger}$ respectively.
By the inclusion principle of Hermitian matrices, (see Theorem
4.3.15 of \cite{Horn}) the smallest eigenvalues of  $\Delta Z_R
\Delta Z_R^{\dagger}$ are larger than the corresponding smallest
eigenvalues of $\Delta Z_S \Delta Z_S^{\dagger}$, i.e.,
\[ \mu_k \ \geq \ \nu_k , \ \ \ 1 \leq k \leq n_t.
\]
Since $||\Delta Z_S||_F^2 \ \dot \leq  \ \text{SNR}$, it follows that
every eigenvalue $\nu_k$ of $\Delta Z_S \Delta Z_S^{\dagger}$ is
bounded above by SNR.  If the ST code ${\cal Z}_S$ is designed to
operate at rate $r \log (\text{SNR})$ bits per channel use, then
we have \bean \det (\Delta Z_R \Delta Z_R^{\dagger}) & \dot \geq &
\frac{\text{SNR}^{T-r}}{\text{SNR}^{T-n_t}} \\
& = & \text{SNR}^{n_t-r} \eean and it follows that the row-deleted
code is also clearly optimal.
\end{proof}

\begin{ex} \label{ex:row_deleted}(Row-Deleted ST Code Derived from CDA)
The CDA-based square ST codes presented in
Section~\ref{sec:square_CDA} constitute an example of clearly
optimal ST code.  From \eqref{eq:LeftRegular}, each codeword $Z$
in the row-deleted $(n_t \times T)$ normalized ST code
${\mathcal{Z}}$ takes on the form
\begin{equation} \label{eq:rect_typeA_matrix}
Z = \theta \ \left[
\begin{array}{cccc}
\ell_0 & \gamma \sigma(\ell_{T-1}) & \cdots & \gamma
\sigma^{T-1}(\ell_1)\\
\ell_1 & \sigma(\ell_{0}) & \cdots & \gamma
\sigma^{T-1}(\ell_2)\\
\vdots & \vdots & \ddots & \vdots\\
\ell_{n_t - 1} & \sigma(\ell_{n_t -2}) & \cdots & \gamma
\sigma^{T-1}(\ell_{n_t})\\
\end{array} \right].
\end{equation}
\end{ex}

\subsection{The Cartesian-Product Construction}
\label{sec:typeB_rect}

\begin{thm} [Cartesian-Product Construction] \label{thm:Cartesian_product}
Let $n_t, T$ be given, $T > n_t$. Let $T$ be partitioned into the
integers $T_k \geq n_t,  \  1 \leq k \leq K,$ satisfying
$\sum_{k=1}^K T_k=T$.  Let $\{{\cal Z}^{(i)}\}_{i=1}^K$ be a
collection of $K$ clearly optimal ST codes of size $(n_t \times
T_k)$.  Then the Cartesian product
\[
{\cal Z} \ = \ \bigotimes_{k=1}^K {\cal Z}^{(k)} \] comprised of
$(n_t \times T)$ code matrices of the form \[ Z \ = \ [Z^{(1)} \
Z^{(2)} \hdots Z^{(K)} ], \ \  \ Z_i \in {\cal Z}^{(i)}
\] is also clearly optimal.
\end{thm}
\begin{proof} It is clear that if each individual code matrix $Z^{(k)} \in {\cal Z}^{(k)}$,
satisfies the energy and rate requirements, the same is true of
the product code.  It remains to verify that the determinant condition is met.  Consider the difference
\bean
\Delta Z & = & [Z_1^{(1)} \ Z_1^{(2)} \hdots Z_1^{(K)}] \ - \ \\
& & [Z_2^{(1)} \ Z_2^{(2)} \hdots Z_2^{(K)}] \\
& = & [\Delta Z^{(1)} \  \Delta Z^{(2)} \hdots \ \Delta Z^{(K)} ]
\eean between any two distinct code matrices in the product code
${\cal Z}$. At least one of the $\Delta Z^{(k)}$, say, $\Delta
Z^{(k_0)}$ must be nonzero. Next, note that we can write \bea
\Delta Z \Delta Z^{\dagger} %& = & %[\Delta Z^{(1)} \  \Delta Z^{(2)} \hdots \ \Delta Z^{(K)} ] \\
%& &
%[\Delta Z^{(1)} \  \Delta Z^{(2)} \hdots \ \Delta Z^{(K)} ]^{\dagger} \\
%& = & \sum_{k=1}^K [\Delta Z^{(k)}] [\Delta Z^{(k)}]^\dagger \\
& = & [\Delta Z^{(k_0)}] [\Delta Z^{(k_0)}]^{\dagger} \\
& & \ \ + \sum_{k=1, k \neq k_0}^K [\Delta Z^{(k)}] [\Delta Z^{(k)}]^\dagger \\
& = & A  + B \eea where $A \ = \ [\Delta Z^{(k_0)}] [\Delta
Z^{(k_0)}]^{\dagger}$ and $B \ = \ \sum_{k=1, k \neq k_0}^K
[\Delta Z^{(k)}] [\Delta Z^{(k)}]^\dagger$. Let \bean
 \mu^{(A)}_1 \leq  \mu^{(A)}_2 \leq   \cdots  \leq \mu^{(A)}_{n_t} & & \text{and} \\
% \mu^{(B)}_1 \leq  \mu^{(B)}_2 \leq  \cdots  \leq \mu^{(B)}_{n_t} \\
  \mu^{(A+B)}_1 \leq  \mu^{(A+B)}_2 \leq   \cdots  \leq \mu^{(A+B)}_{n_t}
 \eean
 denote the ordered eigenvalues of the Hermitian matrices $A,A+B$ respectively.
 Then from a theorem of Weyl (see Theorem 4.3.1 of \cite{Horn}), we have that
 \[
 \mu_i^{(A+B)} \ \geq \  \mu_i^{(A)}, \ \ 1 \leq i \leq n_t.
 \]
 As a result, we have that
 \bean
 \det ([A+B]) & \geq & \det(A) \\
 & \dot \geq & \text{SNR}^{n_t-r}
 \eean
 and thus the determinant condition is also met.
\end{proof}

%\begin{note}
%In situations where the Cartesian-product construction and the
%row-deletion construction are both applicable, while the
%Cartesian-product construction has the advantage of a smaller
%signal constellation thereby leading to reduced implementation
%complexity, the row-deletion construction will likely have the
%advantage of larger coding gain.
%\end{note}

\subsection{Simulation Results}

The simulations in this section are related to the row-deletion
construction of Theorem~\ref{thm:row_deletion}.

They present the performance of a rectangular $n_t=n_r=2, T=3$ ST
code operating at $6$ bpcu alongside that of the Golden
code~\cite{BelRekVit} operating at $4$ and $12$ bpcu. The code
matrices in the $(n_t=2, T=3)$ rectangular ST code are obtained by
deleting the top row in the $( n_t=3 \times T=3)$ CDA-based
perfect~\cite{OggRekBelVit} code. With reference to
Theorem~\ref{thm:suff_condn}, this code is an admissible starting
point for applying the row-deletion construction because as shown
in \cite{EliSetKum}, the parent $( n_t=3 \times T=3)$ perfect ST
code satisfies the sufficiency condition in
Theorem~\ref{thm:suff_condn} relating to D-MG optimality of the
parent square ST code.

\begin{figure}[!h] \begin{center}\includegraphics[width=0.63\columnwidth]{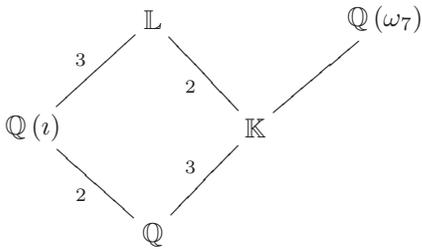} \end{center} \caption{The number
fields relevant to the simulated $(2 \times 3)$ ST code.}
\label{fig:simulated_code}
\end{figure}

We explain the structure of the rectangular code with the aid of
Fig.~\ref{fig:simulated_code}.

%\begin{center}
%\hspace*{0.8in}
%\begin{figure}[!h] \xymatrix{ \ \ \ \ & & {\mathbb L}
%\ar@{-}[rd]_{2} \ar@{-}[ld]_{3} &  & \Q \left( \omega_{7} \right) \ar@{-}[ld] \\
%\ \ \ \ & \Q \left( \imath \right) \ar@{-}[rd]_{2} & & {\mathbb K} \ar@{-}[ld]_{3} & \\
%\ \ \ \ & & \Q &  } \caption{The number fields relevant to the
%simulated $(2 \times 3)$ ST code.} \label{fig:simulated_code}
%\end{figure}
%\end{center}
%

In Fig.~\ref{fig:simulated_code}, $\mathbb{K}$ is the subfield
$\mathbb{Q}(\omega_7+\omega_7^6)$ of $\mathbb{Q}(\omega_7)$.  The
extension $\mathbb{L}/\mathbb{Q}(\imath)$ is cyclic and the
automorphism
\[
\sigma : \sigma(\omega_7) \rightarrow \omega_7^3, \ \
\sigma(\imath)=\imath
\]
is a generator of $\text{Gal} \left( \mathbb{L}/\mathbb{Q}(\imath)
\right) $.    Set $\gamma= \frac{2+\imath}{1+2\imath}$ and
\[ \{ \beta_1, \beta_2, \beta_3 \} \ = \ \{ x, \sigma(x),
\sigma^{2}(x)\} \] where
\[ x = \sum_{k=1}^{2} \tau^{3k}\biggl(
(\omega_7^4)(1-\omega_7)\prod_{k=0}^{2} (1-\omega_7^{3^k})
\biggr),
\]
with $\tau: \omega_7 \rightarrow \omega_7^3$ a generator of
$\text{Gal}\left( \mathbb{Q}(\omega_7) /\mathbb{Q} \right)$. Then
$\{\beta_1, \beta_2, \beta_3\}$ forms an integral basis for ${\cal
O}_{\mathbb{L}}/\mathbb{Q}(\imath)$ and the code matrices in the
$( 3 \times 3)$ ST code are then of the form
\[
X  \ = \  \left[ \begin{array}{ccc}
l_0               &  \gamma\sigma(l_2)        &  \gamma\sigma^2(l_1)\\
l_1               &  \sigma(l_0)        &  \gamma\sigma^2(l_2)\\
l_2               &  \sigma(l_1)        &  \sigma^2(l_0) ,
\end{array} \right]
\]
where
\[
\ell_i \ \in \ \left\{ \sum_{j=1}^3 f_{ij} \beta_j, \ f_{ij} \in
{\cal A}_{\text{QAM}} \right\} \ \subseteq \ {\cal
O}_{\mathbb{L}}.
\]

This particular choice of integral basis along with the property
$| \gamma|=1$ turns out (see \cite{OggRekBelVit},
\cite{EliSetKum}) to ensure that the collection of code matrices,
after vectorization, forms a cubic
constellation~\cite{OggRekBelVit}. Moreover, each antenna element
transmits the same average amount of energy in each time slot.  It
turns out that the resulting ST code not only satisfies the
sufficiency condition identified in Theorem~\ref{thm:suff_condn}
for D-MG optimality, it also has excellent probability-of-error
performance at low-moderate values of SNR.

In our simulation of the rectangular ST code, we have chosen the
QAM constellation ${\cal A}_{\text{QAM}}$ to have size $M^2=4$.
Thus the size of the $( 3 \times 3)$ ST code equals $4^9$.
Deleting the first row does not change the size of the ST code and
hence the rectangular code also has the same size and hence
transmits $6$ bpcu.  From the row-deletion theorem,
Theorem~\ref{thm:row_deletion}, this rectangular code also
satisfies the sufficiency condition for D-MG optimality.  From the
simulation it is seen that the performance of the rectangular code
also tracks channel outage.

\begin{figure}
\begin{center}\includegraphics[angle=-90,width=2.5in]{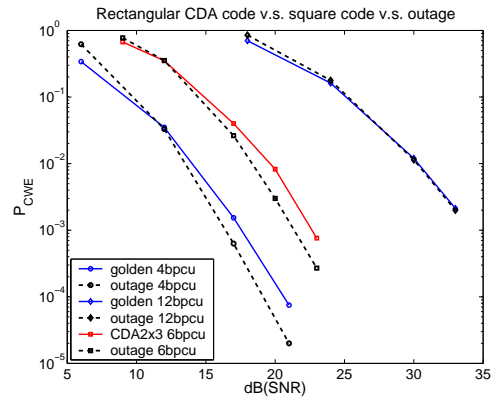}\end{center}
\caption{Simulation of the $(n_t=n_r=2, T=3)$ rectangular code and
comparison with outage at 6 bits per channel use (bpcu). The
performance of the $(n_t=n_r=T=2$) Golden code (along with the
relevant plots of channel outage), is also shown alongside for 4
and 12 bpcu. } \label{fig:simulation}
\end{figure}

\section*{Appendix I: Number Theory Primer
\label{app:numb_theory}}

\setcounter{subsection}{0}

Example references for the relevant number theory include \cite{Lan,Lon,Mar,Rib}.

\subsection{Field Extensions}

Let $\mathbb{E},\mathbb{F}$ be fields such that $\mathbb{F} \subseteq \mathbb{E}$. Then $\mathbb{E}$ is said to
be an extension field of $\mathbb{F}$. $\mathbb{E}$ is naturally a vector space over
$\mathbb{F}$. The extension $\mathbb{E}/\mathbb{F}$ is said to be finite if this dimension is
finite. The degree of a finite extension $\mathbb{E}/\mathbb{F}$ is the dimension of
$\mathbb{E}$ as a vector space over $\mathbb{F}$ and the notation $[\mathbb{E}:\mathbb{F}]$ is used to
denote the degree of the extension. If $\mathbb{E}$ is a finite extension
of $\mathbb{F}$ and $\mathbb{L}$ is a finite extension of $\mathbb{E}$, then we have
\[
[\mathbb{L}:\mathbb{F}] \ = \ [\mathbb{L}:\mathbb{E}][\mathbb{E}:\mathbb{F}] .
\]
A number field is a field that is a finite extension of the field $\mathbb{Q}$ of rational numbers.
Unless otherwise specified, all fields encountered will be assumed to be number fields.
Hence all fields will have characteristic zero and all extensions will be of finite degree.

If $\mathbb{E}$ is an extension of $\mathbb{F}$, then $\alpha \in \mathbb{E}$ is said to be algebraic over $\mathbb{F}$ if $\alpha$
is the zero of some nonzero polynomial $f(x) \in \mathbb{F}[x]$.  $\mathbb{E}$ is said to be an algebraic
extension of $\mathbb{F}$ if every element of $\mathbb{E}$ is algebraic over $\mathbb{F}$.  Every finite extension
is an algebraic extension.

\vspace*{0.2in}

\begin{lem} \label{lem:compositum_of_two_fields}
Let ${\mathbb{S}} = {\mathbb{S}}_1{\mathbb{S}}_2\cdots
{\mathbb{S}}_r$ be the compositum of the fields
${\mathbb{S}}_1$,${\mathbb{S}}_2$,$\hdots$, ${\mathbb{S}}_r$. If
each ${\mathbb{S}}_i$ is an extension of ${{\mathbb{F}}}$ of
degree $m_i$, where the $m_i$ are pairwise relatively prime, then
${\mathbb{S}}$ is an extension over ${{\mathbb{F}}}$ of degree
$\prod_{i=1}^r m_i$.
\end{lem}

\vspace*{0.2in}

\subsection{Algebraic Integers}

An element $\theta$ in a  number field $\mathbb{F}$ is said to be
an algebraic integer if $\theta$ is the zero of a monic polynomial
with rational integer (i.e., elements of $\mathbb{Z}$)
coefficients. The set of all the algebraic integers  in
$\mathbb{F}$ forms a ring called the ring of integers of
$\mathbb{F}$ and is denoted by ${\cal O}_{\mathbb{F}}$.   It is
also referred to as the integral closure of $\mathbb{Z}$ in
$\mathbb{F}$.   If $[\mathbb{\mathbb{L}}:\mathbb{F}]$ is a finite
extension of number fields, then the ring of integers ${\cal
O}_{\mathbb{L}}$ of $\mathbb{L}$ is precisely the collection of
all elements in $\mathbb{L}$ that are the zeros of monic
polynomials with coefficients in ${\cal O}_{\mathbb{F}}$, i.e.,
the integral closure of ${\cal O}_{\mathbb{F}}$ in $\mathbb{L}$ is
${\cal O}_{\mathbb{L}}$.

An integral basis for an extension $\mathbb{L} / \mathbb{F}$ of
number fields is a vector-space basis $\{ \alpha_1, \alpha_2,
\hdots, \alpha_n\}$ for $\mathbb{L}/ \mathbb{F}$ satisfying the
additional requirements that \bit \item $\alpha_i \in {\cal
O}_{\mathbb{L}}$, all $i$ \item Every element in ${\cal
O}_{\mathbb{L}}$ is expressible as a linear combination of
elements in the basis with coefficients lying in ${\cal
O}_{\mathbb{F}}$, i.e., $x \in {\cal O}_{\mathbb{L}}$ implies
\[
x \ = \ \sum_{i=1}^n c_i \alpha_i, \ \ c_i \in {\cal
O}_{\mathbb{F}}.
\]\eit
It is known that every number field extension has an integral
basis.

The term ``rational integer'' is often used to distinguish the
elements of $\mathbb{Z}$ from those in the ring of algebraic
integers of an extension field of $\mathbb{Q}$.

\subsection{Galois Theory}

The Galois group of $\mathbb{E}/\mathbb{F}$ is defined as the set
of all automorphisms $\sigma$ of $\mathbb{E}$ that fix every
element of $\mathbb{F}$, i.e., \bean
\text{Gal}(\mathbb{E}/\mathbb{F}) & = &  \{ \sigma : \mathbb{E} \rightarrow \mathbb{E} \mid \sigma \ \ \text{  is an automorphism of $\mathbb{E}$} \\
& &  \text{ and $\sigma(f) =f, \ \text{all} \ f \in \mathbb{F}$ }
\} . \eean This set forms a group under the composition operator.
The size of the Galois group of the extension
$\mathbb{E}/\mathbb{F}$ is always $\leq [\mathbb{E}:\mathbb{F}]$.
The extension is said to be Galois if equality holds.  An Abelian
(cyclic) extension $\mathbb{E}/\mathbb{F}$ is a Galois extension
in which the Galois group is Abelian (cyclic).

We digress briefly to state a lemma related to cyclic groups.

\vspace*{0.2in}

\begin{lem} \label{lem:quotient_gp_is_cyclic} Let $G$ be a cyclic group of size $g$ and $H$ the unique subgroup of
size $h$ with $h |g$.  Then $G/H$ is cyclic of size $g/h$.
\end{lem}
\begin{proof} Let $\psi$ be the homomorphism from $G$ onto $\psi(G)$ defined by
\[ \psi(x) = x^{h} \ ,\  x \in G \ .\]
Clearly, $\psi (G)$ is a cyclic subgroup of G of order $g/h$. The
kernel $K_\psi$ of $\psi$ is the set of all elements of $G$ whose
order divides $h$.  It follows that the kernel is a cyclic
subgroup of size $h$ and since $H$ is the unique subgroup of this
size, it follows that
\[K_\psi \ = \ H \ . \]
It follows that $G/H \ \cong \ \psi(G)$ and is therefore cyclic.
\end{proof}

\vspace*{0.1in}

A subfield $\mathbb{E} \supset \mathbb{F}$ of $\mathbb{L}$ is said
to be fixed (elementwise) by a subgroup $H$ of $\text{Gal}
(\mathbb{L}/\mathbb{F})$ if
\[
\sigma(x) \ = \ x \ \ \text{for all} \ \ x \in \mathbb{E} \ \ \text{  and all } \ \ \sigma \in H.
\]

\begin{prop} \label{prop:FTGT} (Fundamental Theorem of Galois
Theory) Let ${\mathbb{K}_1}$ be a finite Galois extension of
${{\mathbb{F}}}$. Then there is a one-one correspondence between
the subfields ${\mathbb{K}}$ of ${\mathbb{K}_1}$ containing
${{\mathbb{F}}}$ and subgroups $H$ of
$G=\mbox{\text{Gal}}({\mathbb{K}_1}/{{\mathbb{F}}})$.  The
correspondence maps a subgroup $H$ of $G$ to the largest subfield
${\mathbb{K}}$ of ${\mathbb{K}_1}$ fixed by the subgroup and
subfields ${\mathbb{K}}$ of ${\mathbb{K}_1}$ containing
${{\mathbb{F}}}$ to the largest subgroup of $G$ fixing the
subfield. Moreover, under this correspondence,
 \bit  \item $\text{Gal}(\mathbb{K}_1 / \mathbb{K}) \ = \
H$ \item $\text{Gal}(\mathbb{K} / \mathbb{F}) \ \cong \ G/H$\eit
\end{prop}

\begin{figure}[h!]
\begin{center}
\centerline{\epsfig{figure=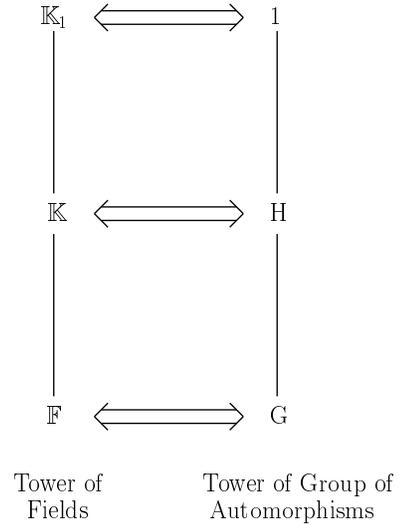,height=70mm}}
%\centerline{\epsfig{angle=-90,figure=arqevents.eps,width=120mm}}
\caption{The Fundamental Theorem of Galois Theory
\label{fig:fundthm}}
\end{center}
\end{figure}

\begin{lem}
\label{lem:compositum_is_cyclic} Let ${\mathbb{S}} =
{\mathbb{S}}_1{\mathbb{S}}_2\cdots {\mathbb{S}}_r$ be the
compositum of the fields
${\mathbb{S}}_1$,${\mathbb{S}}_2$,$\hdots$, ${\mathbb{S}}_r$. If
each ${\mathbb{S}}_i$ is a cyclic Galois extension over
${{\mathbb{F}}}$ of degree $n_i$, where the $n_i$ are pairwise
relatively prime, then ${\mathbb{S}}$ is a cyclic Galois extension
over ${{\mathbb{F}}}$ of degree $\prod_{i=1}^r n_i$.
\end{lem}

\begin{proof} Consider the compositum ${\mathbb{S}}= {\mathbb{S}}_1{\mathbb{S}}_2 \hdots {\mathbb{S}}_r$.
Each ${\mathbb{S}}_i/{{\mathbb{F}}}$ is a cyclic Galois extension
of order $n_i$ such that $(n_i,n_j) = 1 \ \forall \ i\neq j$.
Thus,
\begin{eqnarray*} {\mathbb{S}}_i \cap {\mathbb{S}}_j &=& {{\mathbb{F}}} \ \forall \ i \neq j\\
\Rightarrow \text{Gal}({\mathbb{S}}/{{\mathbb{F}}}) &=&
\text{Gal}({\mathbb{S}}_1/{{\mathbb{F}}}) \times \cdots \times
\text{Gal}({\mathbb{S}}_r/{{\mathbb{F}}}) .
\end{eqnarray*}
It follows that ${\mathbb{S}}/{{\mathbb{F}}}$ is cyclic of degree
$n$.
\end{proof}

\vspace*{0.2in}

\subsection{Prime-Ideal Decomposition in
Number-Field Extensions}

The integral closure ${\cal O}_{\mathbb{F}}$ of $\mathbb{Z}$ in a
number field $\mathbb{F}$ is a Dedekind domain and hence every
prime ideal of ${\cal O}_{\mathbb{F}}$ is a maximal ideal.  To
distinguish primes in $\mathbb{Z}$ from prime ideals of the ring
of integers ${\cal O}_{\mathbb{F}}$ of a number field
$\mathbb{F}$, the term ``rational prime'' is often used to
describe prime elements of $\mathbb{Z}$.   Every ideal $I$ of
${\cal O}_{\mathbb{F}}$ has a unique factorization as the products
of powers of prime ideals. Let $\mathbb{E}$ be a finite Galois
extension of $\mathbb{F}$ and let ${\cal O}_{\mathbb{E}}$ denote
the ring of integers of $\mathbb{E}$.  If ${\frak p}$ is a prime
ideal of ${\cal O}_{\mathbb{F}}$, then the ideal ${\frak p}{\cal
O}_{\mathbb{E}}$ of ${\cal O}_{\mathbb{E}}$ has a unique
factorization of the form \beq {\frak p} \ {\cal O}_{\mathbb{E}} \
= \ \prod_{i=1}^g \beta_i^{e}, \label{eq:p_factorization} \eeq for
distinct prime ideals $\beta_i$ of ${\cal O}_{\mathbb{E}}$.  The
ideal ${\frak p}$ can be recovered from any of the $\beta_i$ via
\[
{\frak p} \ = \ \beta_i \cap {\cal O}_{\mathbb{F}} .
\]
The exponent $e$ is called the ramification index of $\beta_i$
over ${\frak p}$ and written $e(\beta_i/{\frak p})$.  This number
is the same for all $\beta_i$. We will also loosely refer to
$e(\beta_i/{\frak p})$ as the ramification index of ${\frak p}$ or
the ramification index of $\beta_i$.    One may naturally regard
the field ${\cal O}_{\mathbb{E}} / \beta_i$ as an extension of the
field ${\cal O}_{\mathbb{F}} / {\frak p}$ and the degree of this
extension is called the relative degree of $\beta_i$ over ${\frak
p}$ and written $f(\beta_i / {\frak p})$.  This relative degree is
also the same for all $i$.   The decomposition group $G_{\frak p}$
of the ideal ${\frak p}$ of ${\cal O}_{\mathbb{F}}$ is the largest
subgroup of $\text{Gal} (\mathbb{E}/\mathbb{F})$ which fixes
${\frak p}$, i.e.,
\[
G_{\frak p} \ = \ \{ \sigma \in \text{Gal} (\mathbb{E}/\mathbb{F})  \mid  \ \ \sigma(\nu)
\in  {\frak p} \text{  whenever  } \nu \in {\frak p}\}  .
\]
It turns out that the index
\[
\mid \text{Gal} (\mathbb{E}/\mathbb{F}) / G_{\frak p} \mid
\]
of the decomposition group equals the integer $g := g(\beta_i /
{\frak p})$ appearing in the factorization of ${\frak p} {\cal
O}_{\mathbb{E}} $ in \eqref{eq:p_factorization}.  Moreover, \beq
e(\beta_i/{\frak p}) \ f(\beta_i / {\frak p}) \ g(\beta_i / {\frak
p}) \ = \ [\mathbb{E}:\mathbb{F}] \ = n \ .
\label{eq:efg_equals_n} \eeq  The prime ideal ${\frak p}$ is said
to be inert if
\[
f(\beta_i / {\frak p}) \ = \ [\mathbb{E}:\mathbb{F}] \ = n \
\]
in which case, $e(\beta_i / {\frak p})=g(\beta_i / {\frak p})=1$.

Let $\mathbb{L}$ be a Galois extension of $\mathbb{E}$.  If
\[
{\cal O}_{\mathbb{L}} \beta_i \ = \ \prod_{j=1}^{g'} \gamma_j^{e'}
\]
is the unique factorization of the ideal ${\cal O}_{\mathbb{L}}
\beta_i$ in ${\cal O}_{\mathbb{L}}$ in terms of prime ideals
$\gamma_j$, then we have that \bea
 e(\gamma_j/{\frak p}) \ := \ e^{'} & = &  e( \gamma_j / \beta_i )e( \beta_i / {\frak p} ) \label{eq:_eeq_e_1_times_e_2} \\
f(\gamma_j/{\frak p}) & = &  f( \gamma_j / \beta_i )f( \beta_i /
{\frak p} ) \label{eq:f_eq_f_1_times_f_2} \\
g(\gamma_j/{\frak p}) \ := \ g^{'} & = &  g( \gamma_j / \beta_i
)g( \beta_i / {\frak p} )  \label{eq:g_eq_g_1_times_g_2} \eea

\subsection{Cyclotomic Extensions}

Cyclotomic extensions are extensions of the rationals of the form
$\mathbb{Q}(\omega_m) / \mathbb{Q}$ where
\[
\omega_m = \exp \left( \frac{ \imath 2 \pi}{m} \right) \] for some
integer $m \geq 3$.  The degree of this extension equals $\phi
(m)$ where $\phi$ is Euler's totient function.  Such extensions
are of interest here as by a theorem of Kronecker-Weber
\cite{Lon}, every Abelian extension (and therefore every cyclic
extension) of $\mathbb{Q}$ is contained in a cyclotomic extension.

For $m \geq 3$, let ${\mathbb{Z}}_m$ denote the set of integers
modulo $m$ and let ${\mathbb{Z}}^*_m$ be the multiplicative group
formed by the elements $a \in {\mathbb{Z}}_m$ such that $(a,m)=1$.
The following facts will be useful in our study of cyclotomic
extensions:

\vspace*{0.2in}

\begin{lem} \cite{Rib} \label{lem:Pofm}  The Galois
group $\text{Gal}({\mathbb{Q}}(\omega_m)/{\mathbb{Q}}) \cong
{\mathbb{Z}}^*_m $.
\end{lem}

\vspace*{0.2in}

\begin{lem} \cite{Rib} \label{lem:zmstar_is_cyclic}
$ {\mathbb{Z}}^*_m $ is a cyclic group iff $m = 2,4,p^e,2p^e$
where $e\geq 1$ and $p$ is an odd prime.
\end{lem}

\vspace*{0.2in}

\begin{lem} \cite{Rib} \label{lem:zevenstar_is_not_cyclic}
For $e \geq 3$,
\[
\mathbb{Z}^*_{2^e} \ \cong \ \mathbb{Z}_2 \times
\mathbb{Z}_{2^{e-2}}.
\]
\end{lem}

From Lemmas~\ref{lem:Pofm} and \ref{lem:zmstar_is_cyclic} it
follows that:

\vspace*{0.2in}

\begin{cor} \label{cor:cyc_extn_is_cyclic_iff}
Let $m \geq 2$ be an integer. Then the Galois extension
$\mathbb{Q}(\omega_m) / \mathbb{Q}$ is cyclic iff $m =
2,4,p^e,2p^e$.
\end{cor}

\vspace*{0.2in}

\begin{lem} \label{cor:cyc_extn_is_cyclic_iff}
The ring of algebraic integers of a cyclotomic field $\mathbb{Q}(\omega_m)$
is the ring $\mathbb{Z}[\omega_{m}]$, i.e., the algebraic closure
of $\mathbb{Z}$ in $\mathbb{Q}(\omega_m)$ is $\mathbb{Z}[\omega_{m}]$.
\end{lem}

\vspace*{0.2in}

\begin{lem} \label{lem:ram_index_gt_1} Let $p$ be
prime and $m \geq 1$.  The only rational prime $q$ for which the
prime ideal $q \mathbb{Z}[\omega_{p^m}]$ is ramified (i.e., has
ramification index $e >1$ ) in
$\mathbb{Q}(\omega_{p^m})/\mathbb{Q}$ is the prime $p$ itself.
\end{lem}

\vspace*{0.1in}

\begin{lem} \label{lem:decomp_prime_power_cyc_fields} \cite{Rib}
Let $p$ be prime, $m$ an integer $\geq 1$ and let $\omega=\exp
\left( \frac{ \imath 2 \pi }{p^m} \right)$. Let $q$ be any prime
number distinct from $p$, let $f\geq 1$ be the smallest integer
such that $q^{f} \equiv 1 \pmod{ p^{m}}$ and let $g=
\phi(p^{m})/f$. Then $q \mathbb{Z}(\omega_{p^m}) \ = \
\beta_1\cdots\beta_{g}$ where $\beta_1,\cdots,\beta_{g}$ are
distinct prime ideals of $\mathbb{Z}(\omega_{p^m})$.  In
particular, if the order of $q \pmod{p^m} = \phi(p^m)$, then
$f=\phi(p^m)$, $g=1$, so that the prime ideal $q
\mathbb{Z}(\omega_{p^m})$ remains inert in
$\mathbb{Q}(\omega_{p^m}) / \mathbb{Q}$.
\end{lem}

\section*{Appendix II: Replacing the QAM Constellation with a HEX Constellation}

\setcounter{paragraph}{0}
\paragraph{HEX Constellation}

For $M \geq 2$, $M$ even, we define
\[
{\cal A}_{\text{HEX}} \ = \ \left\{ a + \omega_3 b \mid \ |a|, |b|
\leq (M-1), \ a, b \ \ \text{odd}  \right\} ,
\] and
will term the resulting constellation the HEX constellation, (see
\cite{OggRekBelVit}). For any particular value of $M$, a slightly
different collection of points from the lattice $\{ a +\omega_3
b\}$ may be preferable from the point of view of improving the
shaping gain, but for the purposes of constructing codes that are
D-MG optimal, this constellation will suffice. Note that as with
the $M^2$-QAM constellation, we have \bea x \in {\cal
A}_{\text{HEX}} &
\Rightarrow & |x|^2 \leq 2M^2 \ \ \text{  and  } \nonumber \\
\mathbb{E}(|x|^2) & = & \frac{M^2-1}{3} \label{eq:HEX_properties}
\eea assuming that every constellation point is chosen at random.

\paragraph{Endowing the NVD Property}

The discussion in Section~\ref{sec:endowing_nvd} carries over to
the HEX case if one replaces the ring $\mathbb{Z}[\imath]$ with
the ring $\mathbb{Z}[\omega_3]$ of Eisenstein integers since every
nonzero element $x \in \mathbb{Z}[\omega_3]$ has $|x| \geq 1$.

\paragraph{Proof of D-MG Optimality}

It follows from \eqref{eq:HEX_properties} that the proof of
optimality remains unchanged in the case of the HEX constellation.

\paragraph{Constructing D-MG Optimal CDA Codes}

An examination of Section~\ref{sec:square_CDA} shows that the
crucial ingredient needed for constructing a D-MG optimal
CDA-based ST code over the HEX constellation is the construction
of a cyclic field extension $\mathbb{L}/\mathbb{Q}(\omega_3)$ such
that the corresponding ring extension ${\cal
O}_{\mathbb{L}}/\mathbb{Z}[\omega_3]$ contains a prime ideal that
remains inert.

\begin{figure}[!h]
\[ \includegraphics[width=0.73\columnwidth]{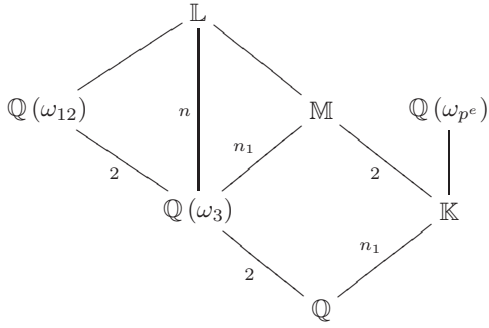} \] \caption{The number
fields that appear in the HEX constellation construction for $n
\neq 0 \pmod{4}$.} \label{fig:constn_hex}
\end{figure}

We provide constructions for the case $n \neq 0 \pmod{4}$.  Let
$p^e$ be the smallest power of an odd prime $p$ such that $n |
\phi(p^e)$.  From Lemma~\ref{lem:k_is_cyclic}, we know that
$\mathbb{Q}(\omega_{p^e})$ contains a subfield $\mathbb{K}$ that
is a cyclic extension of $\mathbb{Q}$ of degree $n_1$. Let
$\mathbb{M}$ be the compositum of $\mathbb{Q}(\omega_3)$ and
$\mathbb{K}$ and $\mathbb{L}$ be the compositum of
$\mathbb{Q}(\omega_{12})$ and $\mathbb{M}$ (see Fig.
~\ref{fig:constn_hex}).

%\begin{figure}[!h]
%\xymatrix{ & {\mathbb L} \ar@{-}[ld] \ar@{-}[dd]_{n}\ar@{-}[rd] & & \\
%\Q \left( \omega_{12} \right) \ar@{-}[rd]_{2}& & {\mathbb M}
%\ar@{-}[rd]_{2} \ar@{-}[ld]_{n_1} &  \Q \left( \omega_{p^{e}} \right) \ar@{-}[d]\\
%& \Q \left( \omega_3 \right) \ar@{-}[rd]_{2} & & {\mathbb K} \ar@{-}[ld]_{n_1}  \\
%& & \Q & } \caption{The number fields that appear in the HEX
%constellation construction for $n \neq 0 \pmod{4}$.}
%\label{fig:constn_hex}
%\end{figure}

\begin{figure}[!h]
\[\includegraphics[width=0.61\columnwidth]{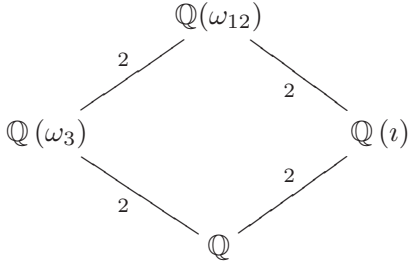}\] \caption{This figure
helps identify an inert prime ideal in the extension
$\mathbb{Q}(\omega_{12})/\mathbb{Q}(\omega_3)$.}
\label{fig:diamond_12}
\end{figure}

Consider the extension
 $\mathbb{Q}(\omega_{12})/\mathbb{Q}(\omega_3)$ (Fig.~\ref{fig:diamond_12}).  This extension is
cyclic of degree $2$ and the Galois group is cyclic having
generator $\sigma: \omega_{12} \rightarrow \omega_{12}^7$.

%\begin{figure}[!h]
%\xymatrix{ & & \mathbb{Q}(\omega_{12})
%\ar@{-}[rd]_{2} \ar@{-}[ld]_{2} &  \\
%& \Q \left( \omega_3 \right) \ar@{-}[rd]_{2} & & \Q \left( \imath \right)  \ar@{-}[ld]_{2}  \\
%& & \Q & } \caption{This figure helps identify an inert prime
%ideal in the extension
%$\mathbb{Q}(\omega_{12})/\mathbb{Q}(\omega_3)$.}
%\label{fig:diamond_12}
%\end{figure}

Let $q$ be a rational prime satisfying \[ q \ = \ \left\{
\begin{array}{rl} \rho & \pmod{p^e}
\\ 1 &
\pmod{3} \\
3 & \pmod{4} . \end{array} \right. \] Then the ideal $q
\mathbb{Z}$ splits into the product of two prime ideals $\beta_1$,
$\beta_2$ in $\mathbb{Q}(\omega_3)/\mathbb{Q}$ since $q = 1
\pmod{3}$. On the other hand, $q \mathbb{Z}$ is inert in
$\mathbb{Q}(\imath)/\mathbb{Q}$ since $q = 3 \pmod{4}$.  Hence it
follows that each of the prime ideals $\beta_i$ is inert in
$\mathbb{Q}(\omega_{12})/\mathbb{Q}(\omega_3)$.  Since $q = \rho
\pmod{p^e}$, it follows that the ideal $q \mathbb{Z}$ is inert in
$\mathbb{K}/\mathbb{Q}$, hence the ideals $\beta_i$ remain inert
in $\mathbb{M}/\mathbb{Q}(\omega_3)$, therefore inert in
$\mathbb{L}/\mathbb{Q}(\omega_{12})$ and as a consequence of their
inertness in $\mathbb{Q}(\omega_{12})/\mathbb{Q}(\omega_3)$, inert
in $\mathbb{L}/\mathbb{Q}(\omega_3)$ as well.

Then $\mathbb{M}/\mathbb{Q}(\omega_3)$ and
$\mathbb{L}/\mathbb{Q}(\omega_3)$ are the desired cyclic
extensions of $\mathbb{Q}(\omega_3)$ for the cases $e_0=0, e_0=1$
respectively.  Note that in either extension, the ideals $\beta_i$
are inert.

\section*{Appendix III: Miscellaneous Proofs}

\subsection*{A.  Proof of Lemma~\ref{lem:det_in_F} -- Determinant Lies in the Center}

\begin{proof} We have
\bea (\sum_{i=0}^{n-1} z^i \ell_i ) \cdot z &= &
(\sum_{i=0}^{n-1}
z^{i+1} \sigma(\ell_i) ) \ , \\
 & = & z (\sum_{i=0}^{n-1}
z^i \sigma(\ell_i) ) \ , \\
\therefore z^{-1}(\sum_{i=0}^{n-1} z^i \ell_i ) \cdot z &= &
\sum_{i=0}^{n-1} z^i\sigma(\ell_i)  . \eea It follows as a result,
that the left-regular representations of
\[
\sum_{i=0}^{n-1} z^i \ell_i  \ \text{and of } \ \ \sum_{i=0}^{n-1}
z^i \sigma(\ell_i)
\]
are similar and therefore have the same determinant.  However,
from inspection of \eqref{eq:LeftRegular}, it follows that the
left-regular representation of $\sum_{i=0}^{n-1} z^i
\sigma(\ell_i)$ equals $\sigma(A)$ where $A$ is the left regular
representation of $\sum_{i=0}^{n-1} z^i \ell_i $.  It follows that
$A$ and $\sigma(A)$ are similar and hence have the same
determinant, i.e., \bean
\det ( \sigma (A) ) & = &  \det (A) \\
\text{i.e., } \ \ \sigma (\det(A)) & = &  \det (A) , \eean so that
$\det (A) \in \mathbb{F}$.
\end{proof}

\subsection*{B.  Proof of \eqref{eq:mismatch_ev_bd} -- Mismatched Eigenvalue Bound }

%\begin{thm} \label{thm:de_min} For a given channel realization $H$ and a
%particular $\Delta X$, the squared Euclidean distance $d_E^2(\Delta X,H) \ = \
%||\theta H \Delta X ||_F^2$ is lower bounded by
%\begin{eqnarray}\label{eq:mismatched_eig} d_E^2(\Delta X,H) \geq \theta^2 \sum_{i=1}^{n_t} \lambda_i
%l_i.
%\end{eqnarray}
%\end{thm}
From \eqref{eq:opens_to_mismatch_ev_bd}, we have \bea
\frac{d_E^2(\Delta X, H)}{\theta^2} & = & \text{Tr} ( H \Delta X
\Delta X^{\dagger} H^{\dagger}) .    \nonumber \eea Using
$\text{Tr}(AB)=\text{Tr}(BA)$ and the eigenvector decompositions
$\Delta X \Delta X^{\dagger}=V L V^{\dagger}$ and
$H^{\dagger}H=U^{\dagger}\Lambda U$, $U,V$ unitary, $\Lambda,L$
diagonal, this can be rewritten in the form, \bea
\frac{d_E^2(\Delta X, H)}{\theta^2} & = &  \text{Tr} ( V^{\dagger} U^{\dagger} \Lambda U V L   )  \nonumber  \\
%& & \ \text{($U,V$ unitary, $\Lambda,L$ diagonal) }    \nonumber  \\
& = &   \sum_{i,j=1}^{n_t} \lambda _i l_j |d_{ij}|^2, \label{eq:d_ij_expn} \\
& & \ \text{($D=[d_{ij}]=UV$, also unitary)} . \eea Let $a_{ij} =
|d_{ij}|^2$. Then,
\begin{eqnarray} \label{eq:constraints}
\sum_i a_{ij} \ = \ \sum_j a_{ij} = 1
\end{eqnarray}
The RHS of \eqref{eq:d_ij_expn} is equal to the sum of the elements  of the matrix
\begin{eqnarray} \label{eq:original}
\left[ \begin{array}{cccc}
  \lambda_1l_1a_{11} & \lambda_1l_2a_{12} & \cdots & \lambda_1l_na_{1{n_t}} \\
  \lambda_2l_1a_{21} & \lambda_2l_2a_{22} & \cdots & \lambda_2l_na_{2{n_t}} \\
  \vdots & \vdots & \ddots & \vdots \\
  \lambda_{n_t}l_1a_{{n_t}1} & \lambda_{n_t}l_2a_{{n_t}2} & \cdots & \lambda_{n_t}l_{n_t}a_{{n_t}{n_t}} \\
\end{array} \right]
\end{eqnarray}
We want to show that setting
\[ a_{ij} = \left\{ \begin{array}{cc}
1,& i=j\\
0,& \text{ otherwise} \end{array} \right., \] minimizes this sum subject to \eqref{eq:constraints}.
Consider any valid assignment of the $\{a_{i,j}\}$. If $a_{11} \neq 1$, then, at least two entries, one each in the first
row and first column are non-zero, say $a_{p1} = \kappa_1$ and $a_{1q} =
\kappa_2$ (where, neither $p$ nor $q$ is $1$). Define $\kappa = \min \{
\kappa_1, \kappa_2 \}$. Consider now the matrix below, obtained by shifting
``weights'' $\kappa$ from $a_{p1}$ and $a_{1q}$ to $a_{11}$ and $a_{pq}$ respectively
while satisfying \eqref{eq:constraints} as shown below. \beq
\begin{array}{c}
 \left[
\begin{array}{cccccc}
  l_1\lambda_1(a_{11} + \kappa) & \cdots & l_q\lambda_1(a_{1q} - \kappa) & \cdots & l_{n_t}\lambda_1a_{1{n_t}} \\
  l_1\lambda_2a_{21} & \cdots & l_q\lambda_2a_{2q} & \cdots & l_{n_t}\lambda_2a_{2{n_t}} \\
  \vdots  & \ddots & \vdots & \ddots & \vdots\\
  l_1\lambda_p(a_{p1} - \kappa) & \cdots & l_q\lambda_p(a_{pq} + \kappa) & \cdots & l_{n_t}\lambda_pa_{p{n_t}} \\
  \vdots & \ddots & \vdots & \ddots & \vdots\\
  l_1\lambda_{n_t}a_{{n_t}1} & \cdots & l_q\lambda_{n_t}a_{{n_t}q} & \cdots & l_{n_t}\lambda_{n_t}a_{{n_t}{n_t}}
\end{array} \right]\\\\
\end{array}
 \label{eq:modified}
\eeq

The difference between the sums of the entries of the matrices in
(\ref{eq:original}) and (\ref{eq:modified}) equals
\begin{eqnarray*}
 & & \ \kappa \lambda_1 l_1 +
\kappa \lambda_p l_q - \kappa \lambda_p l_1 - \kappa \lambda_1
l_q\\
&=& \kappa \underbrace{[\lambda_1 - \lambda_p]}_{positive}
\underbrace{[l_1 - l_q]}_{negative} \text{ is negative}
\end{eqnarray*}
where we have used \eqref{eq:ordering_eigenvalues}.
Thus we see that shifting weights towards the diagonal entry $a_{11}$ results in a reduction in
$d_E^2(\Delta X,H)$. Repeated application of this procedure gives us  $a_{11}=1$
and $a_{1i},a_{i1} = 0$ for $i \neq 1$.  Repeating this procedure with the first row and column removed
leads us to setting $a_{22}=1$ as well and so on,  leading to $a_{ii}=1$, all $i$.
 $\hfill{\square}$

%\subsection*{C.  D-MG Optimal ST Code Construction for the HEX Constellation}
%
%\begin{thm} \label{thm:optimal_CDA}
%Let $T=n_t=n$. Then the square CDA based space-time code
%\begin{equation}
%\mathcal{X} = \left\{\left[
%\begin{array}{cccc}
%  \ell_0 & \gamma \sigma (\ell_{n-1}) &  \hdots & \gamma \sigma^{n-1} (\ell_1) \\
%  \ell_1 & \sigma (\ell_0) &  \hdots & \gamma \sigma^{n-1} (\ell_2) \\
%  \vdots & \vdots & \ddots & \vdots \\
%  \ell_{n-1} & \sigma (\ell_{n-2}) & \hdots  & \sigma^{n-1} (\ell_0) \\
%\end{array}%
%\right] \right\},
%\end{equation}
%where \bit
%    \item the entries $\ell_i$ appearing in the code matrices are
%restricted to belong to the set
%$\mathcal{A}_{\tiny{\text{HEX}}}(\beta_1,\hdots,\beta_n)$ in which
%the set of elements $(\beta_1,\hdots,\beta_n)$ constitute an
%integral basis for $\mathbb{L}/\mathbb{F}$
%    \item $\gamma$ is a non-norm element lying in $\mathbb{Z}[\omega_3]$
%\eit is optimal (after normalization using a suitable scalar
%$\theta$) with respect to the D-MG tradeoff for any number $n_r$
%of receive antennas.
%\end{thm}

\begin{center}
ACKNOWLEDGEMENT
\end{center}

Thanks are due to B. Sethuraman for the particular proof of
Lemma~\ref{lem:det_in_F} appearing in the paper as well as for
drawing our attention to the results on the eigenvalues of
submatrices of Hermitian matrices contained in \cite{Horn}. Thanks
are also due to B. Sundar Rajan and E. Viterbo for making
available preprints of their recent work.

\bibliographystyle{IEEEbib}

\end{document}